\documentclass[letterspace]{article}
\pdfoutput=1

\usepackage{jheppub,setspace}
\usepackage{url,comment}
\usepackage{times}
\usepackage{latexsym}
\usepackage{graphicx, graphics, hyperref, amsmath, amssymb, slashed, color, bbm,amsthm, array}
\usepackage[usenames,dvipsnames]{xcolor}
 \usepackage{subfigure}
\usepackage{mdwlist}
\usepackage{multirow}

\numberwithin{equation}{section}

\def \beg#1{\begin{#1}} 
\def \be{\beg{equation}}
\def \bea{\beg{eqnarray}}
\def \eea{\end{eqnarray}}
\def \ee{\end{equation}}

\def\bc{\begin{center}}
\def\ec{\end{center}}

\def\nn{\nonumber}

\title{Stimuli reduce the dimensionality of cortical activity\linebreak
{\normalsize {\it Front Syst Neurosci} 2016, \normalfont 10:11. doi: 10.3389/fnsys.2016.00011}
}

\author[1]{Luca Mazzucato}
\author[1,2]{Alfredo Fontanini}
\author[1,2]{and Giancarlo La Camera}

\affiliation[1]{Department of Neurobiology and Behavior and}
\affiliation[2]{Graduate Program in Neuroscience, State University of New York at Stony Brook, Stony Brook, NY 11794}
  
\bigskip
\abstract{  
The activity of ensembles of simultaneously recorded neurons can be represented as a set of points in the space of firing rates. Even though the dimension of this space is equal to the ensemble size, neural activity can be effectively localized on smaller subspaces. The dimensionality of the neural space is an important determinant of the computational tasks supported by the neural activity. Here, we investigate the dimensionality of neural ensembles from the sensory cortex of alert rats during periods of ongoing (inter-trial) and stimulus-evoked activity. We find that dimensionality grows linearly with ensemble size, and grows significantly faster during ongoing activity compared to evoked activity. We explain these results using a spiking network model based on a clustered architecture. The model captures the difference in growth rate between ongoing and evoked activity and predicts a characteristic scaling with ensemble size that could be tested in high-density multi-electrode recordings. Moreover, we present a simple theory that predicts the existence of an upper bound on dimensionality. This upper bound is inversely proportional to the amount of pair-wise correlations and, compared to a homogeneous network without clusters, it is larger by a factor equal to the number of clusters. The empirical estimation of such bounds depends on the number and duration of trials and is well predicted by the theory. Together, these results provide a framework to analyze neural dimensionality in alert animals, its behavior under stimulus presentation, and its theoretical dependence on ensemble size, number of clusters, and correlations in spiking network models.
\

\bigskip
\noindent Emails: $\textsf{luca dot mazzucato}$, $\textsf{alfredo dot fontanini}$, $\textsf{giancarlo dot lacamera at stonybrook.edu}$
\noindent 
}


\begin{document}

\setcounter{tocdepth}{2}
\maketitle

\section{Introduction}

Understanding the dynamics of neural activity and how it is generated in cortical circuits is a fundamental question in Neuroscience. The spiking activity of ensembles of simultaneously recorded neurons can be represented in terms of sequences of firing rate vectors, as shown e.g. in frontal \cite{abeles1995cortical,seidemann1996simultaneously,durstewitz2010abrupt}, gustatory \cite{jones2007natural,mazzucato2015dynamics}, motor \cite{kemere2008detecting}, premotor and somatosensory cortex \cite{poncealvarez2012dynamics}. The dimension of each firing rate vector is equal to the number of ensemble neurons $N$ and the collection of rate vectors across trials takes the form of a set of points in the $N$-dimensional space of firing rates. Such points may not fill the whole space, but be restricted to lie inside a lower-dimensional subspace (see e.g. \cite{ganguli2008one}). Roughly, dimensionality is the minimal number of dimensions necessary to provide an accurate description of the neural dynamics. If ensemble neurons are independent of each other, neural activities at different times will scatter around in the space of firing rate, filling a large portion of the space. In this case, dimensionality will be maximal and equal to the size of the ensemble $N$. At the other extreme, if all neurons are strongly correlated, ensemble activity localizes along a line. In this case, dimensionality is minimal and equal to one. These simple examples suggest that dimensionality captures information about the structure of a cortical circuit and the functional relations among the simultaneously recorded neurons, such as their firing rates correlation computed over timescales of hundreds of milliseconds.  

Different definitions of dimensionality have been introduced for different tasks and across neural systems \cite{ganguli2008one,churchland2010cortical,abbott2011interactions,ganguli2012compressed,cadieu2013neural,rigotti2013importance,gao2015simplicity}. Such measures of dimensionality can shed light on the underlying neural computation; for example, they can predict the onset of an error trial in a recall task \cite{rigotti2013importance}, or can allow the comparison of classification accuracy between different brain areas (e.g., IT vs. V4) and synthetic algorithms \cite{cadieu2013neural}. Here, we investigate a measure of dimensionality closely related to the firing rate correlations of simultaneously recorded neurons \cite{abbott2011interactions}; such correlations may provide a signature of feature-based attention \cite{cohen2009attention} and other top-down cognitive factors \cite{nienborg2012decision}. We elucidate the dependence of dimensionality on experimental parameters, such as ensemble size and interval length, and we show that it varies across experimental conditions. We address these issues by comparing recordings of ensembles of neurons from the gustatory cortex (GC) of alerts rats to a biologically plausible network model based on neural clusters with recurrent connectivity. This model captures neural activity in GC during periods of ongoing and stimulus-evoked activity, explaining how the spatiotemporal dynamics of ensemble activity is organized in sequences of metastable states and how single-neuron firing rate distributions are modulated by stimulus presentation \cite{mazzucato2015dynamics}. Here, we show that the same model expounds the observed dependence of dimensionality on ensemble size and how such dependence is reduced by the presentation of a stimulus. By comparing the clustered network model with a homogeneous network without clusters, we find that the clustered network has a larger dimensionality that depends on the number of clusters and the firing rate correlations among ensemble neurons. A simple theory explains these results and allows extrapolating the scaling of dimensionality to very large ensembles.  Our theory shows that recurrent networks with clustered connectivity provide a substrate for high-dimensional neural representations, which may lead to computational advantages. 

\section{Methods}
\label{methods}

\subsection{Experimental procedures}
\label{experimentalprocedures}
Adult female Long Evans rats were used for this study \cite{samuelsen2012effects,mazzucato2015dynamics}. Animals received ad lib. access to food and water, unless otherwise mentioned. Movable bundles of sixteen microwires attached to a ``mini-microdrive" \cite{fontanini2006state,samuelsen2012effects} were implanted in GC (AP 1.4, ML $\pm$ 5 from bregma, DV$-4.5$ from dura). After electrode implantation, intra-oral cannulae (IOC) were inserted bilaterally \cite{fontanini20057,phillips1970rapid}. At the end of the surgery a positioning bolt for restraint was cemented in the acrylic cap. Rats were given at least 7 days for recovery before starting the behavioral procedures outlined below. All experimental procedures were approved by the Institutional Animal Care and Use Committee of Stony Brook University and complied with University, state, and federal regulations on the care and use of laboratory animals. More details can be found in \cite{samuelsen2012effects}.
Rats were habituated to being restrained and to receiving fluids through IOCs, and then trained to self-deliver water by pressing a lever following a $75$ dB auditory cue at a frequency of $4$ KHz. The interval at which lever-pressing delivered water was progressively increased to $40\pm3$ s (ITI). During training and experimental sessions additional tastants were automatically delivered at random times near the middle of the ITI, at random trials and in the absence of the anticipatory cue. Upon termination of each recording session the electrodes were lowered by at least 150 $\mu$m so that a new ensemble could be recorded. A computer-controlled, pressurized, solenoid-based system delivered $\sim40 \mu$l of fluids (opening time $\sim40$ ms) directly into the mouth through a manifold of $4$ polymide tubes slid into the IOC. The following tastants were delivered: $100$ mM NaCl, $100$ mM sucrose, $100$ mM citric acid, and $1$ mM quinine HCl. Water ($\sim50 \mu$l) was delivered to rinse the mouth clean through a second IOC five seconds after the delivery of each tastant. Each tastant was delivered for at least $6$ trials in each condition. Upon termination of each recording session the electrodes were lowered by at least $150 \mu$m so that a new ensemble could be recorded.
Evoked activity periods were defined as the interval after tastant delivery (time $t=0$ in our figures) and before water rinse (time $t=5$ s). Only trials in which the tastants were automatically delivered were considered for the analysis of evoked activity, to minimize the effects of cue-related expectations \cite{samuelsen2012effects}. Ongoing activity periods were defined as the $5$ s-long intervals at the end of each inter-trial period.
The behavioral state of the rat was monitored during the experiment for signs of disengagement. Erratic lever pressing, inconstant mouth movements and fluids dripping from the mouth indicated disengagement and led to the termination of the experiment. In addition, since disengagement from the task is also reflected in the emergence of high power oscillations in local field potentials, occurrences of such periods were removed offline and not analyzed further \cite{fontanini2008behavioral}. 

\subsection{Data analysis}
Single neuron action potentials were amplified, bandpass filtered (at $300-8$ KHz), digitized and recorded to a computer (Plexon, Dallas, TX). Single units of at least $3:1$ signal-to-noise ratio were isolated using a template algorithm, cluster cutting techniques and examination of inter-spike interval plots (Offline Sorter, Plexon, Dallas, TX). All data analyses and model simulations were performed using custom software written in Matlab (Mathworks, Natick, MA, USA), Mathematica (Wolfram Research, Champaign, IL), and C. Starting from a pool of $299$ single neurons in $37$ sessions, neurons with peak firing rate lower than $1$ Hz (defined as silent) were excluded from further analysis, as well as neurons with a large peak around the $6-10$ Hz in the spike power spectrum, which were considered somatosensory \cite{katz2001dynamic,samuelsen2012effects,horst2013reward}. Only ensembles with $3$ or more simultaneously recorded neurons were further analyzed ($167$ non-silent, non-somatosensory neurons from $27$ ensembles). We analyzed ongoing activity in the $5$ seconds interval preceding either the auditory cue or taste delivery, and evoked activity in the $5$ seconds interval following taste delivery in trials without anticipatory cue, wherein significant taste-related information is present \cite{jezzini2013processing}.

\subsection{Hidden Markov Model (HMM) analysis}
\label{hmmanalysis}
Here we briefly outline the procedure used in \cite{mazzucato2015dynamics}, see this reference and \cite{jones2007natural,escola2011hidden,poncealvarez2012dynamics} for further details. Under the HMM, a system of N recorded neurons is assumed to be in one of a predetermined number of hidden (or latent) states \cite{rabiner1989tutorial,rabiner1989tutorial}. Each state $m$ is defined as a vector of $N$ firing rates  $\nu_i (m)$, $i=1,\ldots,N$, one for each simultaneously recorded neuron. In each state, the neurons were assumed to discharge as stationary Poisson processes (Poisson-HMM). We matched the model to the data segmented in $1$-ms bins (see below). In such short bins, we found that typically at most one spike was emitted across all simultaneously recorded neurons. If more than one neuron fired an action potential in a given bin, only one (randomly chosen) was kept for further analysis  (this only occurred in a handful of bins per trial) \cite{escola2011hidden}. We denote by $y_i (t)$  the spiking activity of the $i$-th neuron in the interval $[t,t+dt]$, $y_i (t)=1$ if the neuron emitted a spike and $y_i (t)=0$ otherwise. Denoting with $S_t$ the hidden state of the ensemble at time $t$, the probability of having a spikes from neuron $i$ in a given state $m$ in the interval $[t,t+dt]$ is given by $p(y_i (t)=1|S_t=m)=1-e^{\nu_i (m)dt}$.

The firing rates $\nu_i (m)$ completely define the states and are also called ``emission probabilities" in HMM parlance. The emission and transition probabilities were found by maximization of the log-likelihood of the data given the model via the expectation-maximization (EM), or Baum-Welch, algorithm \cite{rabiner1989tutorial}, a procedure known as ``training the HMM". For each session and type of activity (ongoing vs. evoked), ensemble spiking activity from all trials was binned at $1$ ms intervals prior to training assuming a fixed number of hidden states $M$ \cite{jones2007natural,escola2011hidden}. For each given number of states $M$, the Baum-Welch algorithm was run $5$ times, each time with random initial conditions for the transition and emission probabilities. The range of hidden states $M$ for the HMM analyses were $M_{min}=10$ and $M_{max}=20$ for spontaneous activity, and $M_{min}=10$ and $M_{max}=40$ for evoked activity. Such numbers were based on extensive exploration of the parameter space and previous studies \cite{jones2007natural,miller2010stochastic,escola2011hidden,poncealvarez2012dynamics,mazzucato2015dynamics}. For evoked activity, each HMM was trained on all four tastes simultaneously. Of the models thus obtained, the one with largest total likelihood $M^*$ was taken as the best HMM match to the data, and then used to estimate the probability of the states given the model and the observations in each bin of each trial (a procedure known as ``decoding"). During decoding, only those hidden states with probability exceeding $80\%$ in at least $50$ consecutive bins were retained (henceforth denoted simply as ``states"). State durations were approximately exponentially distributed with median duration $0.60$ s ($95\%$ CIs: $0.07-4.70$) during ongoing activity and $0.30$ s $(0.06-2.80)$ during evoked activity \cite{mazzucato2015dynamics}.
The firing rate fits $\nu_i (m)$ in each trial were obtained from the analytical solution of the maximization step of the Baum-Welch algorithm, 
\be
\label{eqone}
\nu_i (m)=-{1\over dt} \ln\left(1-{\sum_{t=1}^T r_m(t)y_i(t) \over \sum_{t=1}^T r_m(t)}\right)  \, .
\ee
Here, $[y_i (1),\ldots,y_i (T)]$ is the spike train of the $i$-th neuron in the current trial, and $T$ is the total duration of the trial. $r_m (t)=P(S_t=m|y(1),\ldots,y(T))$ is the probability that the hidden state $S_t$ at time $t$ is $m$, given the observations.

\subsection{Dimensionality measure}
\label{dimensionalitymeasure}
We defined the dimensionality of the neural activity as \cite{abbott2011interactions}
\be\label{eqtwo}
d={1\over\sum_{i=1}^N\tilde\lambda_i^2}\ ,
\ee
where the $\tilde\lambda_i$ are the principal eigenvalues expressed as fractions of the total amount of variance explained, i.e. $\tilde\lambda_i=\lambda_i/ (\sum_j\lambda_j )$, where $\lambda_j$ are the eigenvalues of the covariance matrix of the firing rates (see below). 
The dimensionality can be computed exactly in some relevant special cases. The calculation is simplified by the observation that Eq. (\ref{eqtwo}) is equivalent to 
\be
d={[\textrm{Tr}(C_f)]^2\over \textrm{Tr}(C_f^2)} \  ,      
\ee					      				       
where $C_f$ is the true covariance matrix of the firing rate vectors, $\textrm{Tr}(A)=\sum_{i=1}^NA_{{ii}}$ is the trace of matrix $A$, and $\textrm{Tr}(A^2)=\sum_{i,j=1}^NA_{{ij}}A_{ji}$. We consider in the following only the case of firing rates in equal bins, hence we can replace $C_f$ with the covariance matrix of the spike counts $C$ in the definition of $d$:
\be\label{eqthree}
d={[\textrm{Tr}(C)]^2\over Tr(C^2)}={b_N^2\over c_N+a_N }  \ ,
\ee
where for later convenience we have introduced the notation
\be
a_N=\sum_{i=1}^NC_{{ii}}^2 \  ,\qquad 
b_N=\sum_{i=1}^NC_{{ii}} \ , \label{eqfour}\qquad  
c_N=\sum_{i\neq j}^NC_{{ij}}C_{ji} \ .
\ee
Note that $d$ does not depend on the distribution of firing rates, but only on their covariance, up to a common scaling factor.

\noindent {\it Dimensionality in the case of uniform pair-wise correlations}. When all the pair-wise correlations $r_{ij}$  are identical, $r_{ij}=\rho$ for all $i\neq j$, 
\be\label{eqfive}
r_{{ij}}=
\left[
\begin{array}{cccc}
1 & \rho& \cdot & \rho \\
\rho & 1 &  & \cdot \\
\cdot & & \cdot & \rho \\
\rho & \rho & \rho & 1 
\end{array}
\right]  \ ,
\ee
we have $C_{ij}=\rho\sqrt{\sigma _i^2 \sigma _j^2 }$ for $i\neq j$, where $\sigma _i^2=C_{{ii}}$ is the spike count variance. In this case, we find from Eq. (\ref{eqfour}) that
\be
a_N=\sum_{i=1}^N\sigma_i^4 \  ,\qquad
b_N=\sum_{i=1}^N\sigma_i^2 \ , \label{eqsix}\qquad
c_N=\rho^2 (b_N^2-a_N) \ .
\ee
and the dimensionality, Eq. (\ref{eqthree}), is given by
\be\label{eqseven}
d={1\over \rho^2+(1-\rho^2 )g(N)} \ ,    
\ee    			     
where 
$$
g_N={a_N\over b_N^2 }={\sum_{i=1}^N\sigma _i^4 \over (\sum_{i=1}^N\sigma _i^2 )^2} \ ,
$$
Note that since both $a_N$ and $b_N$ scale as $N$ when $N$ is large, in general $g_N\sim {1\over N}$  for large $N$. 
If all spike counts have equal variance, $\sigma _i=\sigma$ , we find exactly $g_N={1\over N}$:
\be\label{eqeight}
d={1\over \rho^2+{1-\rho^2 \over N}}={N\over N\rho^2+1-\rho^2} \ ,
\ee
and the dependence of $d$ on the variance drops out. Note that for uncorrelated spike counts ($\rho=0$) this formula gives $d=N$, whereas for any finite correlation we find the upper bound $d={1\over \rho^2}$. For $N>1$, the dimensionality is inversely related to the amount of pair-wise correlation $\rho$.

\noindent Consider the case where spike counts have variances $\sigma _i^2$ drawn from a probability distribution with mean $\textrm{E}[\sigma _i^2 ]=\sigma^2$ and variance $\textrm{Var}[\sigma _i^2 ]=\delta\sigma^4$, and the pair-wise correlation coefficients $r_{ij}$, for $i\neq j$, are drawn from a distribution with mean $\textrm{E}[r_{ij} ]=\rho$ and variance $\textrm{Var}[r_{ij} ]=\delta\rho^2$. In such a case one can evaluate Eq. (\ref{eqthree}) approximately by its Taylor expansion around the mean values of the quantities in Eqs. (\ref{eqfour}). At leading order in $N$ one 
\be
\textrm{E}[d]\simeq {\textrm{E}[b_N^2 ]\over \textrm{E}[c_N ]+\textrm{E}[a_N ] }={N\sigma^4+\delta\sigma^4\over (N-1) \sigma ^4 (\rho^2+\delta\rho^2)+\sigma ^4+\delta\sigma^4 } \ ,                               \label{eqnine}
\ee
where $\textrm{E}[.]$ denotes expectation. To obtain this result we have used the definitions in Eq. (\ref{eqfour}), from which   
\bea
\textrm{E}[a_N]&=&N(\sigma ^4+\delta\sigma^4) \  ,\nn\\
\textrm{E}[b_N^2]&=&N^2 \sigma ^4+N\delta\sigma^4 \ , \label{eqten}\\   
\textrm{E}[c_N ]&=&(N^2-N)\sigma^4 (\rho^2+\delta\rho^2) \ .\nn
\eea
and the fact that, given a random vector $X_i$ with mean $\mu_i$ and covariance $C_{ij}$, and a constant symmetric matrix $A_{ij}$, the expectation value of the quadratic form $\sum_{i,j}X_iA_{ij} X_j$ is
\be\label{eqeleven}
\textrm{E}[\sum_{i,j}X_i A_{ij} X_j]=\sum_{ij}(A_{ij} C_{ji}+\mu_i A_{ij} \mu_j)\ .  
\ee
In the case of uncorrelated spike counts ($\rho=0,\,\delta\rho=0$), dimensionality still depends linearly on the ensemble size $N$, but with a smaller slope $\sigma^4/(\sigma ^4+\delta\sigma^4 )<1$ compared to the case of equal variances (Eq. (\ref{eqeight}) with $\rho=0$).

\noindent {\it Dimensionality in the case of neural clusters}. Given an ensemble of $N$ neurons arranged in $Q$ clusters (motivated by the model network described later in Sec.~\ref{spikingmodel}), we created ensembles of uncorrelated spike trains for $N\leq Q$ and correlated within each cluster for $N>Q$. Thus, if $N\leq Q$ the correlation matrix is the $N \times N$ identity matrix. If $N>Q$, the $(Q+1)$-th neuron was added to the first cluster, with correlation $\rho$ with the other neuron of the cluster, and uncorrelated to the neurons in the remaining clusters. The $(Q+2)$-th neuron was added to the second cluster, with correlation $\rho$ with the other neuron of the second cluster, and uncorrelated to the neurons in the remaining clusters, and so on. Similarly, the $(2Q+p)$-th neuron $(p\leq Q)$ was added to the $p$-th  cluster, with pair-wise correlation $\rho$ with the other neurons of the same cluster, but no correlation with the neurons in the remaining clusters; and so on. In general, for $N=mQ+p$ neurons (where $m=[N/Q]_-\geq1$ is the largest integer smaller than $N/Q$), the procedure picked $m+1$ neurons per cluster for the first $p$ cluster and $m$ neurons per cluster for the remaining $Q-p$ clusters, with uniform pair-wise correlations $\rho$ in the same cluster while neurons from different clusters were uncorrelated. The resulting correlation matrix $r$ was block diagonal
\be\nn
r=\textrm{diag}(R_1,\ldots,R_Q ) \ ,
\ee
where each of the $Q$ blocks contains the correlations of neurons from the same cluster. Inside each block $R_i$, the off-diagonal terms are equal to the uniform within-cluster correlation $\rho$:
\be\nn
R_i=
\left[
\begin{array}{cccc}
1 & \rho& \cdot & \rho \\
\rho & 1 &  & \cdot \\
\cdot & & \cdot & \rho \\
\rho & \rho & \rho & 1 
\end{array}
\right]  \ ,
\ee
The first $p$ blocks have size $(m+1)\times(m+1)$ and the last $Q-p$ blocks have size $m\times m$, so that $(m+1)p+m(Q-p)=N$. The remaining elements of matrix $r$ (representing pair-wise correlations of neurons belonging to different clusters) were all zero. Recalling that $C_{{ij}}=r_{{ij}} \sigma_i \sigma_j$, one finds $\textrm{Tr}(C)=p b_{m+1}+(Q-p) b_m$ and $\textrm{Tr}(C^2 )=\rho^2 [p b_{m+1}^2+(Q-p) b_m^2]+(1-\rho^2)[p a_{m+1}+(Q-p) a_m]$, where $a_n$ and $b_n$ are defined in Eq.~(\ref{eqsix}), from which one obtains
\be\label{eqtwelve}
d=
\left\{
\begin{array}{cc}
{b_N^2/a_N} \ ,                                    &N\leq Q \\
{[p b_{m+1}+(Q-p) b_m ]^2\over \rho^2 [p b_{m+1}^2+(Q-p) b_m^2]+(1-\rho^2)[p a_{m+1}+(Q-p) a_m]} \ ,& N>Q
\end{array}
\right.
\ee
In the approximation where all neurons have the same variance this simplifies to 
\be\label{eqthirteen}
d=\left\{\begin{array}{cc}
N \ ,                                    &N\leq Q \\
{N\over 1+m\rho^2[1-(Q-p)/N]} \ ,& N>Q
\end{array}
\right.
\ee
Recall that in the formulae above $m$ and $p$ depend on $N$. For finite $\rho$, Eq.~(\ref{eqthirteen}) predicts the bound $d\leq Q/\rho^2$ for any $N$, with this value reached asymptotically for large $N$. When single neuron variances $\sigma_i^2$ are drawn from a distribution with mean $E[\sigma_i^2 ]=\sigma^2$ and variance $\textrm{Var}[\sigma_i^2 ]=\delta\sigma^4$, an expression for the dimensionality can be easily obtained from Eq. (\ref{eqtwelve}) at leading order around the expectation values in Eq. \ref{eqfour} (not shown), with a procedure similar to that use to obtain Eq. (\ref{eqnine}).		

\subsection{Pair-wise correlations}
Given neuron $i$ and neuron $j$'s spike trains, we computed the spike count correlation coefficient $r_{{ij}}$, 
\be
\nn
r_{{ij}}={S_{{ij}}\over \sqrt{S_{{ii}} S_{jj} }} \ ,
\ee
where $S$ is the sample covariance matrix of the spike counts estimated as
\be\label{eqfourteen}
S_{ij}={1\over N_b N_T-1} \sum_{b,s=1}^{N_b,N_T}\left(n_i (b,s)-\langle n_i\rangle \right)\left(n_j (b,s)-\langle n_j\rangle\right) \  ,
\ee
where $n_i (b,s)$ is the spike count of neuron $i$ in bin $b$ and trial $s$. The sum goes over all $N_b$  bins and over all $N_T$ trials in a session, whereas $\langle n_i\rangle$ is the average across trials and bins for neuron $i$. In the main text and figures we present results obtained with a bin size of $200$ ms, but have performed the same analyses with bin sizes varying from $10$ ms to $5$ seconds (see Results for details). 

Significance of the correlation was estimated as follows \cite{renart2010asynchronous}: $N_{shuffle}=200$ trial-shuffled correlation coefficients $r'_{{ij}}$ were computed, then a $p$-value was determined as the fraction of shuffled coefficients $r'_{{ij}}$ whose absolute value exceeded the absolute value of the experimental correlation, $p=(\#(|r'_{{ij}} |>|r_{{ij}}|))/N_{shuffle}$. For example, a correlation $r$ was significant at $p=0.05$ confidence level if no more than $10$ shuffled correlation coefficients out of $200$ exceeded $r$.
The pair-wise correlations of firing rates vectors computed in bins of fixed duration $T$ were given by Eq. (\ref{eqfourteen}) with $n_i (b,s)$ replaced by $n_i (b,s)/T$. Instead, correlations of firing rates vectors inside hidden states (which have variable duration) were estimated after replacing $n_i (b,s)$ in Eq. (\ref{eqfourteen}) with $\nu_i (m,s)$, the firing rate of neuron $i$ in state $m$ in trial $s$. For each trial $s$, this quantity was computed according to Eq. (\ref{eqone}).

\subsection{Estimation of dimensionality}
The eigenvalues $\lambda_j$ in Eq. (\ref{eqtwo}) were found with a standard Principal Component Analysis (PCA) of the set of all firing rate vectors \cite{chapin1999principal}. The firing rate vectors were obtained via the HMM analysis (see Eq. (\ref{eqone})); all data from either ongoing or evoked activity were used. For the analysis of Fig. \ref{figthree}E, where the duration and number of trials were varied, only the firing rate vectors of the HMM states present in the given trial snippet were used (even if present for only a few ms). When firing rate vectors in hidden states were not available (mainly, in ``shuffled" datasets and in asynchronous homogeneous networks, see below for details), the firing rates were computed as spike counts in $T=200$ ms bins divided by $T$, $n_i (b,s)/T$, where $n_i (b,s)$ is as defined in Eq. (\ref{eqfourteen}) (Fig. \ref{figthree}F, \ref{figthree}G, \ref{figsix}E, \ref{figseven}D and \ref{fignine}A). Dimensionality values were averaged across $20$ simulated sessions for each ensemble size $N$; in each session, $40$ trials of $5$ s duration, resulting in $N_T=1,000$ bins, were used (using bin widths of $50$ to $500$ ms did not change the results). Note that for the purpose of computing the dimensionality (Eq. (\ref{figthree})), it is equivalent to use either the binned firing rate $n_i (b,s)/T$ or the spike count $n_i (b,s)$.

In our data, $d$ roughly corresponded to the number of principal components explaining between $80$ to $90\%$ of the variance. However, note that all eigenvalues are retained in our definition of dimensionality given in Eq. (\ref{eqtwo}) above. 

\noindent {\it Shuffled datasets}. The dimensionality of the data as a function of ensemble size $N$ was validated against surrogate datasets constructed by shuffling neurons across different sessions while matching the empirical distribution of ensemble sizes. Comparison analyses between empirical and shuffled ensembles were trial-matched using the minimal number of trials per condition across ensembles, and then tested for significant difference with the Mann-Whitney test on samples obtained from $20$ bootstrapped ensembles. Neurons whose firing rate variance exceeded the population average by two standard deviations were excluded ($8/167$ of non-silent, non-somatosensory neurons).

\noindent {\it Dependence on the number of trials: simulations (Fig. \ref{figseven}E, \ref{figeight}A)}. The estimate of d from data depends on the number and duration of the trials (Fig. \ref{figthree}E and Eq. (\ref{eqsixteen}) below). To investigate this phenomenon in a simple numerical setting we generated $N \times N_T$ ``nominal" firing rates, thought of as originating from $N$ neurons, each sampled $N_T$ times (trials). The single firing rates were sampled according to a log-normal distribution with equal means and covariance leading to Eq. (\ref{eqseven}), i.e., $C_{ij}=\rho\sigma_i \sigma_j (1-\delta_{ij})+\sigma_i^2 \delta_{ij}$, with $\delta_{ij}=1$ if $i=j$, and zero otherwise (note that the actual distribution used is immaterial since the dimensionality only depends on the covariance matrix, see Eq. (\ref{eqthree})). We considered the two cases of equal variance for all ensemble neurons, $\sigma_i=\sigma$ for all $i$ (Fig. \ref{figeight}A) or variances $\sigma_i$ sampled from a log-normal distribution (Fig. \ref{figeight}A and ``+'' in Fig. \ref{figseven}E). The same $N$ and $N_T$ as used for the analysis of the model simulations in Fig. \ref{figseven}D were used (where the ``trials" were $N_T$ bins of $200$ ms in $40$ intervals of $5$ second duration for each ensemble size $N$). The covariance of the data thus generated was estimated according to Eq. (\ref{eqfourteen}), based on which the dimensionality Eq. (\ref{eqthree}) was computed. The estimated dimensionality depends on $N$ and $N_T$ and was averaged across $100$ values of $d$, each obtained as explained above. Note that in this simplified setting increasing the duration of each trial is equivalent to adding more trials, i.e., the effect of having a trial $400$ ms long producing $2$ firing rates (one for each $200$ ms bin) is equivalent to having two trials of $200$ ms duration. In the general case, the effect of trial duration on $d$ will depend on how trial duration affects the variance and correlations of the firing rates. 

\noindent {\it Dependence on the number of trials: theory}. The dependence of dimensionality on the number of trials can be computed analytically under the assumption that $N$ ensemble neurons generate spike counts $n_i$, for $i=1,\ldots,N$, distributed according to a multivariate Gaussian. Since we are interested in the spike-count covariance Eq. (\ref{eqfourteen}), we can assume the spike-count distribution to have zero mean and true covariance $C_{ij}$. The matrix $M=(N_T-1)\cdot S^{(N_T) }$, where $S^{(N_T)}$ is the covariance matrix Eq. (\ref{eqfourteen}) sampled from $N_T$  trials, is distributed according to a Wishart distribution $W_N (C_{ij},N_T-1)$ with $N_T-1$ degrees of freedom \cite{mardia1979multivariate}. Since the variance of the Wishart distribution, 
$$
\textrm{Var}(M_{ij} )=(N_T-1)(C_{ij}^2+C_{ii} C_{jj}) \ ,
$$
is proportional to $N_T$, we obtain the variance of the entries of the sample covariance as
\be\label{eqfifteen}
\textrm{Var}(S_{ij}^{(N_T)} )={1\over N_T-1}(C_{ij}^2+C_{ii} C_{jj}) \ ,
\ee
to be used in the estimator of $d$ (from Eq. (\ref{eqthree}))
$$
\hat{d}={[\textrm{Tr}(S)]^2\over\textrm{Tr}(S^2)}={\widehat{b_N^2 }\over\widehat{c_N}+\widehat{a_N } } ,
$$
where  $\widehat{a_N},\widehat{c_N},\widehat{b_N^2}$ are given by Eq. (\ref{eqfour}) with $C$ replaced by $S$. With a calculation similar to that used to obtain Eq. (\ref{eqnine}), to leading order in $N$ and $N_T$ one finds 
$$
\textrm{E}[\hat{d}]\simeq {\textrm{E}[\widehat{b_N^2}]\over \textrm{E}[\widehat{c_N}]+\textrm{E}[\widehat{a_N}]} \ ,
$$
with
\bea
\textrm{E}[\widehat{a_N}]&=&N(\sigma^4+\delta\sigma^4 )+{2N\sigma^4\over N_T-1} \ ,\nn\\
\textrm{E}[\widehat{b_N^2}]&=&N^2 \sigma^4+N\delta\sigma^4+{2N\sigma^4\over N_T-1} \ ,\nn\\
\textrm{E}[\widehat{c_N} ]&=&(N^2-N) (\rho^2+\delta\rho^2)\sigma^4+(N^2-N)  {1+\rho^2+\delta\rho^2\over N_T-1} \sigma^4\ ,\nn   
\eea     
where we also used Eqs. (\ref{eqten}) and (\ref{eqeleven}), with $\textrm{Var}[\sigma_i^2 ]=\delta\sigma^4$ and $\textrm{Var}[r_{ij} ]=\delta\rho^2$, for $i\neq j$.  In conclusion, at leading order in $N$ and $N_T$ one finds
\be\label{eqsixteen}
            \textrm{E}[\hat{d}]={\left(N+{2\over N_T-1}\right) \sigma^4+\delta\sigma^4\over (N-1)\left(\rho^2+\delta\rho^2+{1+\rho^2+\delta\rho^2\over N_T-1}\right) \sigma^4+\left(1+{2\over N_T-1}\right) \sigma^4+\delta\sigma^4 } \ .  
\ee

\noindent {\it Model fitting}. The dependence of the dataÕs dimensionality on ensemble size $N$ was fitted by a straight line via standard least-squares,
$$
d=\beta_1\cdot N+\beta_0 \ ,
$$
separately for ongoing and evoked activity (Fig. \ref{figthree}B-D and \ref{figsix}B-D). Comparison between the dimensionality of evoked and ongoing activity was carried out with a 2-way ANOVA with condition (evoked vs. ongoing) and ensemble size ($N$) as factors. Since $d$ depends on the number and duration of the trials used to estimate the covariance matrix (Fig. \ref{figthree}E and Eq. (\ref{eqsixteen})), we matched both the number of trials and trial length in comparisons of ongoing and evoked dimensionality. If multiple tastes were used, the evoked trials were each matched to a random subset of an equal number of ongoing trials. 

The dependence of dimensionality d on ensemble size $N$ in a surrogate dataset of Poisson spike trains with mean pairwise correlation $\rho$ (generated according to the algorithm described in the next section) was modeled as Eq. (\ref{eqsixteen}) with $\delta\rho^2=\alpha\rho^2$ and $\delta\sigma^4=\sigma^4=\beta$ (Fig. \ref{figseven}D, dashed lines); $N_T$ was fixed to $1000$ ($40$ trials of $5$ seconds each, segmented in $200$ ms bins). The parameters $\alpha,\beta$ were tuned to fit all Poisson trains simultaneously on datasets with $N=5,10,\ldots,100$ and $\rho= 0,0.01,0.05,0.1,0.2$, with $20$ ensembles for each value (Fig. \ref{figseven}D; only the fits for $\rho= 0,0.1,0.2$ are shown). A standard non-linear least-squares procedure was used \cite{holland1977robust}.

\subsection{Generation of correlated Poisson spike trains}
Ensembles of independent and correlated Poisson spike trains were generated for the analysis of Fig.~(\ref{figseven}). Ensembles of independent stationary Poisson spike trains with given firing rates $\nu_i$ were generated by producing their interspike intervals according to an exponential distribution with parameter $\nu_i$. Stationary Poisson spike trains with fixed pairwise correlations (but no temporal correlations) were generated according to the method reported in \cite{macke2009generating}, that we briefly outline below. 
We split each trial into $1$ ms bins and consider the associated binary random variable $X_i (t)=1$ if the $i$-th neuron emitted a spike in the $t$-th bin, and $X_i (t)=0$ if no spike was emitted. These samples were obtained by first drawing a sample from an auxiliary $N$-dimensional Gaussian random variable $U\sim N(\gamma,\Lambda)$ and then thresholding it into $0$ and $1$: $X_i=1$ if $U_i>0$, and $X_i=0$ otherwise. Here, $\gamma=\{\gamma_1,\gamma_2,\ldots,\gamma_N\}$ is the mean vector and $\Lambda=\{\Lambda_{ij}\}$ is the covariance matrix of the $N$-dimensional Gaussian variable $U$. For appropriately chosen parameters $\gamma_i$ and $\Lambda_{{ij}}$ the method generates correlated spike trains with the desired firing rates $\nu_i$ and pairwise spike count correlation coefficients $r_{{ij}}$. 

The prescription for $\gamma_i$ and $\Lambda_{{ij}}$ is most easily expressed as a function of the desired probabilities $\mu_i$ of having a spike in a bin of width $dt$, $\mu_i=P(X_i (t)=1)$, and the pairwise covariance $S_{{ij}}$ of the random binary vectors $X_i (t)$ and $X_j (t)$, from which $\gamma_i$ and $\Lambda_{{ij}}$ can be obtained by inverting the following relationships:  
\bea
\mu_i&=&\Phi(\gamma_i ) \ ,\nn\\
c_{{ii}}&=&\Phi(\gamma_i )\Phi(-\gamma_i ) \ ,\nn\\
c_{{ij}}&=&\Phi_2 (\gamma_i,\gamma_j,\Lambda_{{ij}} )-\Phi(\gamma_i )\Phi(\gamma_j ) \ ,     \quad i\neq j \ .\nn
\eea
Here, $\Phi(x)$ is the cumulative distribution of a univariate Gaussian with mean $0$ and variance $1$ evaluated at $x$, and $\Phi_2 (x,y,\Lambda)$ is the cumulative distribution of a bivariate Gaussian with means $0$, variances $1$ and covariance $\Lambda$ evaluated at $(x, y)$ (note that the distributions $\Phi$ and $\Phi_2$ are unrelated to the $N$-dimensional Gaussian $U\sim N(\gamma,\Lambda)$). Without loss of generality we imposed unit variances for $U_i$, i.e. $\Lambda_{{ii}}=1$. 

We related the spike probabilities $\mu_i$ to the firing rates $\nu_i$ as $\mu_i=1-e^{-\nu_i dt}$, with $1-\mu_i$ being the probability of no spikes in the same bin. When $dt$ approaches zero, $\mu_i\simeq\nu_i dt$ and the spike trains generated as vectors of binary random variables by sampling $U\sim N(\gamma,\Lambda)$ will approximate Poisson spike trains ($dt=1$ ms bins were used). In order to have a fair comparison with the data generated by the spiking network model (described in the next section), the mean firing rates of the Poisson spike trains were matched to the average firing rates obtained from the simulated data.  
Since $\gamma$ and $\Lambda$ were the same in all bins, values of $X_i (t)$ and $X_i (s)$ were independent for $t\neq s$ (i.e., the spike trains had no temporal correlations). As a consequence, the random binary vectors have the same pair-wise correlations as the spike counts, and the $c_{{ij}}$ are related to the desired $r_{{ij}}$ by $c_{{ij}}=r_{{ij}} \sqrt{\mu_i (1-\mu_i ) \mu_j (1-\mu_j ) }$, where $\mu_i (1-\mu_i )$ is the variance of $X_i$. See \cite{macke2009generating} for further details.

\subsection{Spiking network model}
\label{spikingmodel}
We modeled the data with a recurrent spiking network of $N=5000$ randomly connected leaky integrate-and-fire (LIF) neurons, of which $4000$ excitatory ($E$) and $1000$ inhibitory ($I$). Connection probability $p_{\beta\alpha}$ from neurons in population $\alpha\in E,I$ to neurons in population $\beta\in E,I$ were $p_{EE}=0.2$ and $p_{EI}=p_{IE}=p_{{ii}}=0.5$; a fraction $f=0.9$ of excitatory neurons were arranged into $Q$ different clusters, with the remaining neurons belonging to an unstructured (``background") population \cite{amit1997model}. Synaptic weights $J_{\beta\alpha}$ from neurons in population $\alpha\in E,I$ to neurons in population $\beta\in E,I$  scaled with $N$ as $J_{\beta\alpha}=j_{\beta\alpha}/\sqrt{N}$, with $j_{\beta\alpha}$ constants having the following values (units of mV): $j_{EI}=3.18,j_{IE}=1.06,j_{{ii}}=4.24,j_{EE}=1.77$. Within an excitatory cluster synaptic weights were potentiated, i.e. they took average values of $\langle J\rangle_+=J_+ j_{EE}$ with $J_+>1$, while synaptic weights between units belonging to different clusters were depressed to average values $\langle J\rangle_-=J_- j_{EE}$, with $J_-=1-\gamma f(J_+-1)<1$, with $\gamma=0.5$. The latter relationship between $J_+$ and $J_-$ helps to maintain balance between overall potentiation and depression in the network \cite{amit1997model}. 

Below spike threshold, the membrane potential V of each LIF neuron evolved according to 
\be
\tau_m  {dV\over dt}=-V+\tau_m (I_{\textrm{rec}}+I_\textrm{ext}+I_\textrm{stim} ) \  ,
\nn
\ee
with a membrane time constant $\tau_m=20$ ms for excitatory and $10$ ms for inhibitory units. The input current was the sum of a recurrent input $I_{rec}$, an external current $I_{ext}$ representing an ongoing afferent input from other areas, and an external stimulus $I_{stim}$ representing e.g. a delivered taste during evoked activity only. In our units, a membrane capacitance of $1$ nF is set to $1$. A spike was said to be emitted when $V$ crossed a threshold $V_{thr}$, after which $V$ was reset to a potential $V_{reset}=0$ for a refractory period of $\tau_{ref}=5$ ms. Spike thresholds were chosen so that, in the unstructured network (i.e., with $J_+=J_-=1$), the $E$ and $I$ populations had average firing rates of $3$ and $5$ spikes/s, respectively  \cite{amit1997model}. The recurrent synaptic input $I_{rec}^i$ to unit $i$ evolved according to the dynamical equation
\be\nn
\tau_s  {dI_\textrm{rec}^i\over dt}=-I_\textrm{rec}^i+\sum_{j=1}^NJ_{{ij}}  \sum_k\delta(t-t_k^j )  \  ,
\ee
where $t_k^j$ was the arrival time of $k$-th spike from the $j$-th pre-synaptic unit, and $\tau_s$ was the synaptic time constant ($3$ and $2$ ms for $E$ and $I$ units, respectively), resulting in an exponential post-synaptic current in response to a single spike,  ${J_{{ij}}\over \tau_s}  e^{-t/\tau_s}\Theta(t)$, where $\Theta(t)=1$ for $t\geq0$, and $\Theta(t)=0$ otherwise. The ongoing external current to a neuron in population $\alpha$ was constant and given by
\be\nn
I_\textrm{ext}=N_\textrm{ext} p_{\alpha0} J_{\alpha0} \nu_\textrm{ext} \ ,
\ee
where $N_{ext}=n_E N$, $p_{\alpha0}=p_{EE}$, $J_{\alpha0}=j_{\alpha0}/\sqrt{N}$ with $j_{E0}=0.3$, $j_{I0}=0.1$, and $\nu_{ext}=7$ spikes/s. During evoked activity, stimulus-selective units received an additional input representing one of the four incoming stimuli. The stimuli targeted combinations of neurons as observed in the data. Specifically, the fractions of neurons responsive to $n=1,2,3$ or all $4$ stimuli were $17\% (27/162)$, $22\% (36/162)$, $26\% (42/162)$, and $35\% (57/162)$ \cite{jezzini2013processing,mazzucato2015dynamics}. Each stimulus had constant amplitude $\nu_{stim}$ ranging from $0$ to $0.5 \nu_{ext}$. In the following we measure the stimulus amplitude as percentage of $\nu_{ext}$ (e.g., ``$10\%$" corresponds to $\nu_{stim}=0.1 \nu_{ext}$). The onset of each stimulus was always $t=0$, the time of taste delivery. The stimulus current to a unit in population $\alpha$ was constant and given by 
$$
I_{stim}=N_{ext} p_{\alpha0} J_{\alpha0} \nu_{stim} \ .
$$

\subsection{Mean field analysis of the model}
The stationary states of the spiking network model in the limit of large $N$ were found with a mean field analysis \cite{amit1997model,brunel1999fast,fusi1999collective,curti2004mean,mazzucato2015dynamics}. Under typical conditions, each neuron of the network receives a large number of small post-synaptic currents (PSCs) per integration time constant. In such a case, the dynamics of the network can be analyzed under the diffusion approximation within the population density approach. The network has $\alpha=1,\ldots,Q+2$ sub-populations, where the first $Q$ indices label the $Q$ excitatory clusters, $\alpha=Q+1$ labels the ``background" units, and $\alpha=Q+2$ labels the homogeneous inhibitory population. In the diffusion approximation \cite{tuckwell1988introduction,lansky1999stochastic,richardson2004effects}, the input to each neuron is completely characterized by the infinitesimal mean $\mu_\alpha$   and variance $\sigma_\alpha^2$ of the post-synaptic potential (see \cite{mazzucato2015dynamics} for the expressions of the infinitesimal mean and variance for all subpopulations). 

Parameters were chosen so that the network with $J_+=J_-=1$ (where all $E\to E$ synaptic weights are equal) would operate in the balanced asynchronous regime \cite{van1996chaos,vreeswijk1998chaotic,renart2010asynchronous}, where incoming contributions from excitatory and inhibitory inputs balance out, neurons fire irregular spike trains with weak pair-wise correlations.  

The unstructured network has only one dynamical state, i.e., a stationary point of activity where all $E$ and $I$ neurons have constant firing rate $\nu_E$ and $\nu_I$, respectively. In the structured network (where $J_+>1$), the network undergoes continuous transitions among a repertoire of states, as shown in the main text. To avoid confusion between network activity states and HMM states, we refer to the former as network ``configurations" instead of states. Admissible networks configurations must satisfy the $Q+2$ self-consistent mean field equations \cite{amit1997model}
\be \nn
\nu_\alpha=F_\alpha \left(\mu_\alpha (\overrightarrow{\nu}),\sigma_\alpha^2 (\overrightarrow{\nu})\right)  \ ,
\ee
where $\overrightarrow{\nu}=[\nu_1,\ldots,\nu_Q,\nu_E^{(bg)},\nu_I ]$ is the firing rate vector and $F_\alpha (\mu_\alpha,\sigma_\alpha^2 )$ is the current-to-rate response function of the LIF neurons. For fast synaptic times, i.e. ${\tau_s\over \tau_m} <<1$, $F_\alpha (\mu_\alpha,\sigma_\alpha^2 )$ is well approximated by \cite{Brunel1998,Fourcaud2002}
\be\nn
F_\alpha (\mu_\alpha,\sigma_\alpha )=\left(\tau_\textrm{ref}+\tau_{m,\alpha}\sqrt{\pi}\int_{H_{\textrm{eff},\alpha}}^{\Theta_{\textrm{eff},\alpha}}e^{u^2}  [1+\textrm{erf}(u)]\right)^{-1},
\ee
where 
\bea
\Theta_{\textrm{eff},\alpha}&=&{V_{\textrm{thr},\alpha}-\mu_\alpha\over \sigma_\alpha} +ak_\alpha \ ,\nn\\
H_{\textrm{eff},\alpha}&=&{V_{\textrm{reset},\alpha}-\mu_\alpha\over \sigma_\alpha} +ak_\alpha,\nn
\eea
where $k_\alpha=\sqrt{\tau_{s,\alpha}/\tau_{m,\alpha} }$ is the square root of the ratio of synaptic time constant to membrane time constant, and $a=|\zeta(1/2)|/\sqrt{2}\sim1.03$. This theoretical response function has been fitted successfully to the firing rate of neocortical neurons in the presence of in vivo-like fluctuations \cite{Rauch2003,LaCamera2006,LaCamera2008,giugliano2004single}. 

The fixed points $\overrightarrow{\nu}^*$ of the mean field equations were found with NewtonÕs method \cite{Press2007}. The fixed points can be either stable (attractors) or unstable depending on the eigenvalues $\lambda_\alpha$ of the stability matrix
\be
S_{\alpha\beta}={1\over \tau_{s,\alpha}}\left({\partial F_\alpha \left(\mu_\alpha (\overrightarrow{\nu}),\sigma_\alpha^2 (\overrightarrow{\nu})\right)\over \partial\nu_\beta }-{\partial  F_\alpha \left(\mu_\alpha (\overrightarrow{\nu}),\sigma_\alpha^2 (\overrightarrow{\nu})\right)\over \partial\sigma_\alpha^2 }  {\partial\sigma_\alpha^2\over \partial\nu_\beta }-\delta_{\alpha\beta} \right) \  ,
\nn
\ee
evaluated at the fixed point $\overrightarrow{\nu}^*$ \cite{Mascaro1999}. If all eigenvalues have negative real part, the fixed point is stable (attractor). If at least one eigenvalue has positive real part, the fixed point is unstable. Stability is meant with respect to an approximate linearized dynamics of the mean and variance of the input current:
\bea
\tau_{s,\alpha}  {dm_\alpha\over dt}&=&- m_\alpha+\mu_\alpha (\overrightarrow{\nu}) \ ,\nn\\
{\tau_{s,\alpha}\over 2}  {ds_\alpha^2\over dt}&=&- s_\alpha^2+\sigma_\alpha^2 (\overrightarrow{\nu}) \ , \nn\\
\nu_\alpha (t)&=&F_\alpha \left(m_\alpha (\overrightarrow{\nu}),s_\alpha^2 (\overrightarrow{\nu})\right) \ , \nn
\eea
where $\mu_\alpha$ and $\sigma_\alpha^2$ are the stationary values for fixed $\overrightarrow{\nu}$ given earlier. For fast synaptic dynamics in the asynchronous balanced regime, these rate dynamics are in very good agreement with simulations (\cite{LaCamera2004} Ð see \cite{Renart2004,Giugliano2008} for more detailed discussions). 

\subsection{Metastable configurations in the network model}
The stable configurations of a network with an infinite number of neurons were obtained in the mean field approximation of the previous section and are shown in Fig.~\ref{figfour}B for $Q=30$ and a range of values of the relative potentiation parameter $J_+$. Above the critical point $J_+=4.2$, stable configurations characterized by a finite number of active clusters emerge (grey lines; the number of active clusters is reported next to each line). For a given $J_+$, the firing rate is the same in all active clusters and is inversely proportional to the total number of active clusters. Stable patterns of firing rates are also found in the inhibitory population (red lines), in the inactive clusters (having low firing rates; grey dashed lines), and in the unstructured excitatory population (dashed blue lines). For a fixed value of $J_+$, multiple stable configurations coexist with different numbers of active clusters. For example, for $J_+=5.3$, stable configurations with up to $7$ active clusters are stable, each configuration with different firing rates. This generates multistable firing rates in single neurons, i.e., the property, also observed in the data, that single neurons can attain more than $2$ firing rates across states \cite{mazzucato2015dynamics}. Note that if $J_+\leq5.15$ an alternative stable configuration of the network with all clusters inactive (firing rates $< 10$ spikes/s) is also possible (single brown line).
Strictly speaking, the configurations in Fig.~\ref{figfour}B are stable only in a network containing an infinite number of uncorrelated neurons. In a finite network (or when neurons are strongly correlated) these configurations can lose stability due to strong fluctuations which ignite transitions among the different configurations. Full details are reported in \cite{mazzucato2015dynamics}.

\subsection{Model simulations and analysis of simulated data}
The dynamical equations of the LIF neurons were integrated with the Euler algorithm with a time step of $dt=0.1$ ms. We simulated $20$ different networks (referred to as ``sessions" in the following) during both ongoing and evoked activity. We chose four different stimuli per session during evoked activity (to mimic taste delivery). Trials were $5$ seconds long. The HMM analyses for Figs. \ref{figtwo} and \ref{figfive} were performed on ensembles of randomly selected excitatory neurons with the same procedure used for the data (see previous section ``Hidden Markov Model (HMM) analysis"). The ensemble sizes were chosen so as to match the empirical ensemble sizes ($3$ to $9$ randomly selected neurons). For the analysis of Fig.~\ref{figeight}B, random ensembles of increasing size (from $5$ to $100$ neurons) were used from simulations with $Q=30$ clusters. When the ensemble size was less than the number of clusters ($N\leq Q$), each neuron was selected randomly from a different cluster; when ensemble size was larger than the number of clusters, one neuron was added to each cluster until all clusters were represented, and so on until all $N$ neurons had been chosen. To allow comparison with surrogate Poisson spike trains, the dimensionality of the simulated data was computed from the firing rate vectors in $T=200$ ms bins as explained in Sec. \ref{dimensionalitymeasure}. For control, the dimensionality was also computed from the firing rate vectors in hidden states obtained from an HMM analysis, obtaining qualitatively similar results.

\section{Results}
\label{results}

\begin{figure}[ht]
\begin{center}
\vspace*{-1cm}                                                           
\hspace*{-0.5cm}                                                           
\includegraphics[width=0.9\textwidth]{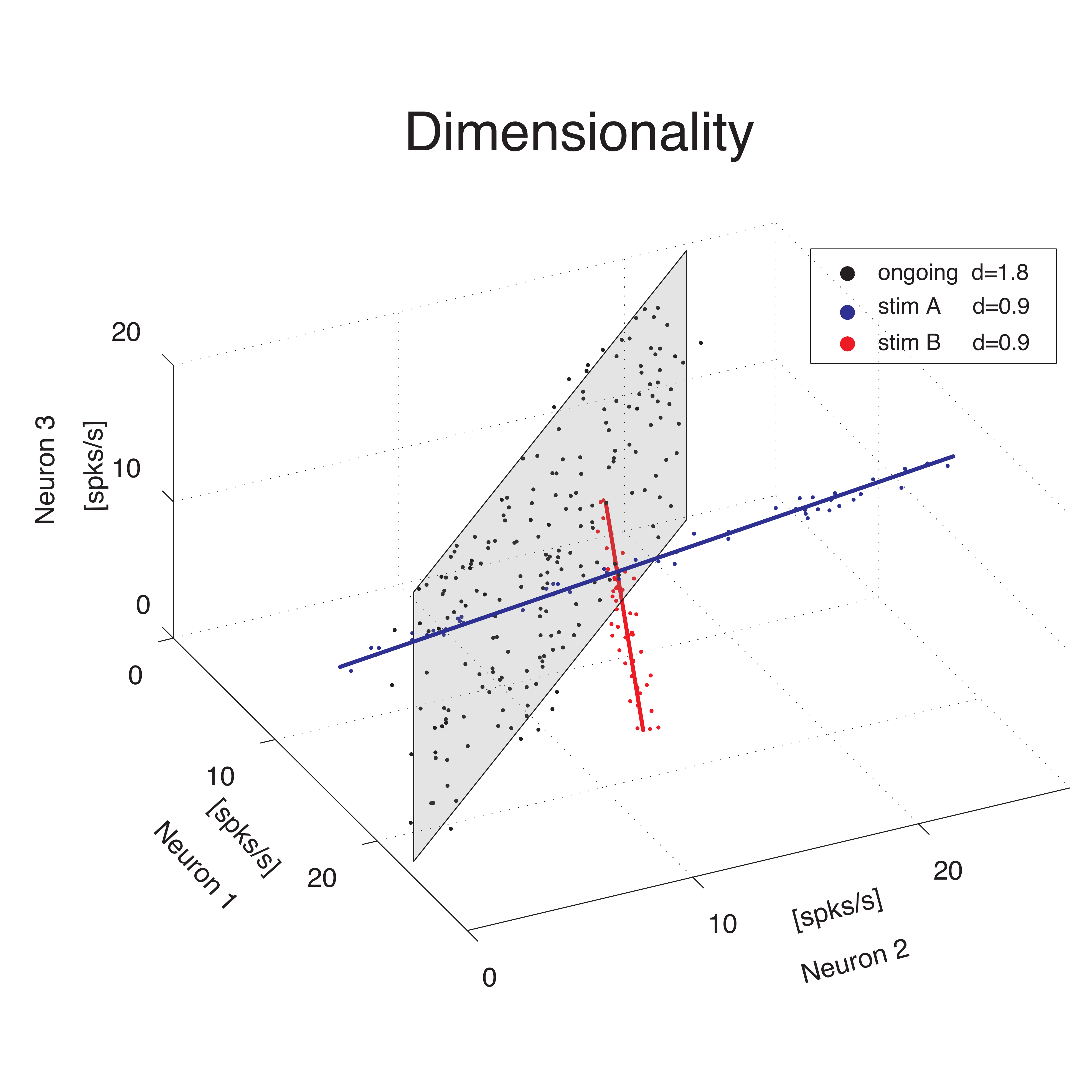}
\end{center}
\vspace*{-1.5cm}
\caption{{\it Dimensionality of the neural representation}. Pictorial representation of the firing rate activity of an ensemble of $N=3$ neurons. Each dot represents a three-dimensional vector of ensemble firing rates in one trial. Ensemble ongoing activity localizes around a plane (black dots cloud surrounding the shaded black plane), yielding a dimensionality of $d=1.8$. Activity evoked by each of two different stimuli localizes around a line (red and blue dots clouds and lines), yielding a dimensionality of $d=0.9$ in both cases.}
\label{figone}
\end{figure}

\subsection{Dimensionality of the neural activity}
We investigate the dimensionality of sequences of firing rate vectors generated in the GC of alert rats during periods of ongoing or evoked activity (see Section~ \ref{experimentalprocedures}). To provide an intuitive picture of the meaning of dimensionality adopted in this paper, consider the firing rate vectors from $N$ simultaneously recorded neurons. These vectors can occupy, a priori, the entire $N$-dimensional vector space minimally required to describe the population activity of $N$ independent neurons. However, the sequence of firing rate vectors generated by the neural dynamics may occupy a subspace that is spanned by a smaller number $m<N$ of coordinate axes. For example, the data obtained by the ensemble of three simulated spike counts in Fig.~\ref{figone} mostly lie on a 2D space, the plane shaded in gray. Although $3$ coordinates are still required to specify all data points, a reduced representation of the data, such as that obtained from PCA, would quantify the dimension of the relevant subspace as being close to $2$. To quantify this fact we use the following definition of dimensionality \cite{abbott2011interactions}
\be
\label{eqthreeresults}
d=\left(\sum_{i=1}^N\tilde \lambda_i^2 \right)^{-1} \ ,
\ee
where $N$ is the ensemble size and $\tilde\lambda_i$ are the normalized eigenvalues of the covariance matrix, each expressing the fraction of the variance explained by the corresponding principal component (see Section~\ref{dimensionalitymeasure} for details). According to this formula, if the first n eigenvalues express each a fraction $1/n$ of the variance while the remaining eigenvalues vanish, the dimensionality is $d=n$. In less symmetric situations, $d$ reflects roughly the dimension of the linear subspace explaining most variance about all data points. In the example of the data on the gray plane of Fig.~\ref{figone}, $d=1.8$, which is close to $2$, as expected. Similarly, data points lying mostly along the blue and red straight lines in Fig.~\ref{figone} have a dimensionality of $0.9$, close to $1$. In all cases, $d>0$ and $d\leq N$, where $N$ is the ensemble size.
 
The blue and red data points in Fig.~\ref{figone} were obtained from a fictitious scenario where neuron $1$ and neuron $2$ were selective to surrogate stimuli A and B, respectively, and are meant to mimic two possible evoked responses. The subspace containing responses to both stimuli A and B would have a dimensionality $d_{A+B}=1.7$, similar to the dimensionality of the data points distributed on the grey plane (meant instead to represent spike counts during ongoing activity in the same fictitious scenario). Thus, a dimensionality close to $2$ could originate from different patterns of activity, such as occupying a plane or two straight lines. Other and more complex scenarios are, of course, possible. In general, the dimensionality will reflect existing functional relationships among ensemble neurons (such as pair-wise correlations) as well as the response properties of the same neurons to external stimuli.  The pictorial example of Fig.~\ref{figone} caricatures a stimulus-induced reduction of dimensionality, as found in the activity of simultaneously recorded neurons from the GC of alert rats, as we show next.

\begin{figure}[t]
\begin{center}
\hspace*{-0.35cm}                                                           
\includegraphics[width=1.02\textwidth]{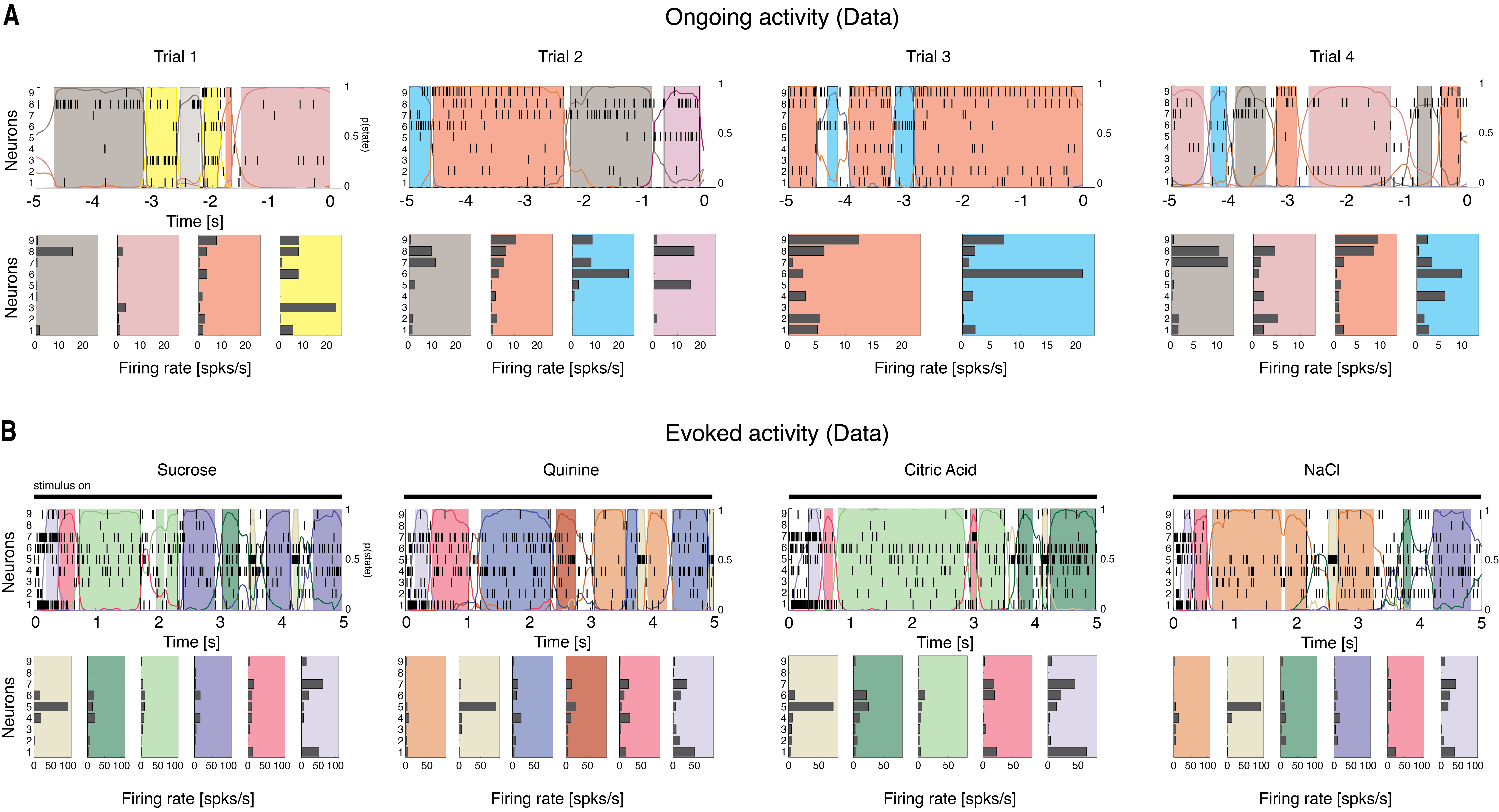}
\end{center}
\vspace*{-0.5cm}
\caption{{\it Ensemble neural activity is characterized by sequences of states}. A: Upper panels: Representative trials from one ensemble of nine simultaneously recorded neurons during ongoing activity, segmented according to their ensemble states (HMM analysis, thin black vertical lines are action potentials; states are color-coded; smooth colored lines represent the probability for each state; shaded colored areas indicate intervals where the probability of a state exceeds $80\%$). Lower panels: Average firing rates across simultaneously recorded neurons (states are color-coded as in the upper panels). In total, 6 hidden states were found in this example session. X-axis for population rasters: time preceding the next event at ($0 =$ stimulus delivery); Y-axis for population rasters: left, ensemble neuron index, right, probability of HMM states; X-axis for average firing rates panels: firing rates (spks/s); Y-axis for firing rate panels: ensemble neuron index. B: Ensemble rasters and firing rates during evoked activity for four different tastes delivered at $t=0$ (the black line on top of the raster plot represents the ``stimulus-on'' period): sucrose, sodium chloride, citric acid and quinine (notations as in panel A). In total, eight hidden states were found in this session during evoked activity.}
\label{figtwo}
\end{figure}

\subsection{Dimensionality is proportional to ensemble size }

\begin{figure}[t]
\vspace*{-2cm}                                                           
\begin{center}
\hspace*{-1.3cm}                                                           
\includegraphics[width=1.13\textwidth]{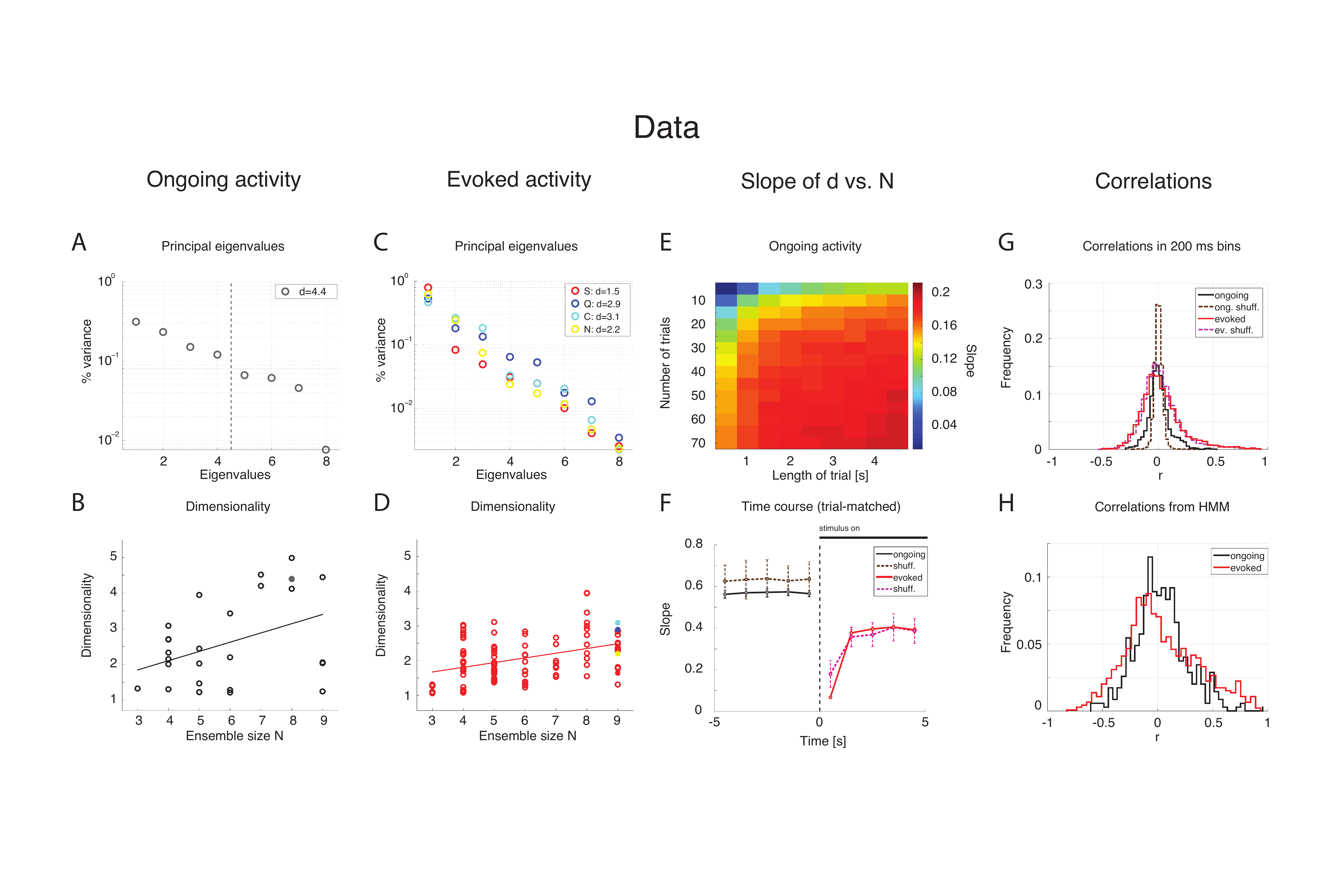}
\end{center}
\vspace*{-2cm}
\caption{{\it Dependence of dimensionality on ensemble size (data)}. A: Fraction of variance explained by each principal eigenvalue for an ensemble of 8 neurons during ongoing activity (corresponding to the filled dot in panel B) in the empirical dataset. The dashed vertical line represents the value of the dimensionality for this ensemble ($d=4.4$). X-axis: eigenvalue number; Y-axis: fraction of variance explained by each eigenvalue. B: Dimensionality of neural activity across all ensembles in the empirical dataset during ongoing activity  (circles, linear regression fit, $d=b\cdot N+a$, $b=0.26\pm 0.12$, $a=1.07\pm 0.74$, $r=0.4$), estimated from HMM firing rate fits on all ongoing trials in each session (varying from $73$ to $129$). X-axis: ensemble size; Y-axis: dimensionality. C: Fraction of variance explained by each principal eigenvalue for the ensemble in panel A during evoked activity. Principal eigenvalues for the four tastes sucrose (S, orange), sodium chloride (N, yellow), citric acid (C, cyan), and quinine (Q, blue) are presented (corresponding to the color-coded dots in panel D). X-axis: eigenvalue number; Y-axis: percentage of variance explained by each eigenvalue. D: Dimensionality of neural activity across all ensembles in the empirical dataset during evoked activity  (notations as in panel B, linear regression: $d=b\cdot N+a$, $b=0.13\pm 0.03$, $a=1.27\pm 0.19$, $r=0.39$), estimated from HMM firing rate fits on evoked trials in each condition (varying from $7$ to $11$ trials across sessions for each tastant). E: The slope of the linear regression of dimensionality ($d$) vs. ensemble size ($N$) as a function of the length of the trial interval and the number of trials used to estimate the dimensionality. X-axis: length of trial interval [s]; Y-axis: number of trials. F: Time course of the trial-matched slopes of $d$ vs. $N$, evaluated with $200$ ms bins in consecutive $1$ s intervals during ongoing (black curve, $t<0$) and evoked periods (red curve, $t>0$; error bars represent SD). A significant time course is triggered by stimulus presentation (see Results for details). The slopes of the empirical dataset (thick curves) were smaller than the slope of the shuffled dataset (dashed curves). X-axis: time from stimulus onset at $t=0$ [s]; Y-axis: slope of $d$ vs. $N$. G: Distribution of pair-wise correlations in simultaneously recorded ensembles (black and red histograms for ongoing and evoked activity, respectively) and shuffled ensembles (brown and pink dashed histograms for ongoing and evoked activity, respectively) from $200$ ms bins. X-axis: correlation; Y-axis: frequency. H: Distribution of pair-wise correlations from HMM states during ongoing (black) and evoked activity (red) for all simultaneously recorded pairs of neurons. X-axis: correlation; Y-axis: frequency.}
\label{figthree}
\end{figure}

We computed the dimensionality of the neural activity of ensembles of $3$ to $9$ simultaneously recorded neurons in the gustatory cortex of alert rats during the $5$ s inter-trial period preceding (ongoing activity) and following (evoked activity) the delivery of a taste stimulus  (said to occur at time $t=0$; see Methods). Ensemble activity in single trials during both ongoing (Fig.~\ref{figtwo}A) and evoked activity (Fig.~\ref{figtwo}B) could be characterized in terms of sequences of metastable states, where each state is defined as a collection of firing rates across simultaneously recorded neurons \cite{jones2007natural,mazzucato2015dynamics}. Transitions between consecutive states were detected via a Hidden Markov Model (HMM) analysis, which provides the probability that the network is in a certain state at every $1$ ms bin (Fig.~\ref{figtwo}, color-coded lines superimposed to raster plots). The ensemble of spike trains was considered to be in a given state if the posterior probability of being in that state exceeded $80\%$ in at least $50$ consecutive 1-ms bins (Fig.~\ref{figtwo}, color-coded shaded areas). Transitions among states were triggered by the co-modulation of a variable number of ensemble neurons and occurred at seemingly random times \cite{mazzucato2015dynamics}. For this reason, the dimensionality of the neural activity was computed based on the firing rate vectors in each HMM state (one firing rate vector per state per trial; see Methods for details).
The average dimensionality of ongoing activity across sessions was $d_{ongoing}=2.6\pm1.2$ (mean$\pm$SD; range: $[1.2, 5.0]$; $27$ sessions). An example of the eigenvalues for a representative ensemble of eight neurons is shown in Fig. \ref{figthree}A, where $d=4.42$.  The dimensionality of ongoing activity was approximately linearly related to ensemble size (Fig. \ref{figthree}B, linear regression, $r=0.4$, slope $b_{ongoing}=0.26\pm 0.12$, $p=0.04$). During evoked activity dimensionality did not differ across stimuli (one-way ANOVA, no significant difference across tastants, $p>0.8$), hence all evoked data points were combined for further analysis. An example of the eigenvalue distribution of the ensemble in Fig. \ref{figtwo}B is shown in Fig. \ref{figthree}C, where $d_{evoked}=1.3 \sim 1.7$ across $4$ different taste stimuli. Across all sessions, dimensionality was overall smaller ($d_{evoked}=2.0\pm 0.6$, mean$\pm$SD, range: $[1.1,3.9]$) and had a reduced slope as a function of $N$ compared to ongoing activity (Fig. \ref{figthree}D, linear regression, $r=0.39$, slope $b_{evoked}=0.13\pm 0.03$, $p<10^{-4}$). However, since dimensionality depends on the number and duration of the trials used for its estimation (Fig. \ref{figthree}E), a proper comparison requires matching trial number and duration for each data point, as described next.

\subsection{Stimulus-induced reduction of dimensionality}

We matched the number and duration of the trials for each data point and ran a two-way ANOVA with condition (ongoing vs. evoked) and ensemble size as factors. Both the main dimensionality ($F_{1,202}=11.93$, $p<0.001$) and the slope were significantly smaller during evoked activity (test of interaction, $F_{6,202}=5.09$, $p<10^{-4}$). There was also a significant effect of ensemble size ($F_{6,202}=18.72$, $p<10^{-14}$), confirming the results obtained with the separate regression analyses. These results suggest that stimuli induce a reduction of the effective space visited by the firing rate vector during evoked activity. This was confirmed by a paired sample analysis of the individual dimensionalities across all $27\times 4=108$ ensembles ($27$ ensemble times $4$ gustatory stimuli; $p<0.002$, Wilcoxon signed-rank test).

\subsection{Dimensionality is larger in ensembles of independent neurons}
 
The dimensionality depends on the pair-wise correlations of simultaneously recorded neurons. Shuffling neurons across ensembles would destroy the correlations (beyond those expected by chance), and would give a measure of how different the dimensionality of our datasets would be compared to sets of independent neurons. We measured the dimensionality of surrogate datasets obtained by shuffling neurons across sessions; because shuffling destroys the structure of the hidden states, firing rates in bins of fixed duration ($200$ ms) were used to estimate the dimensionality (see Methods for details). As expected, the slope of $d$ vs. $N$ was larger in the shuffled datasets compared to the simultaneously recorded ensembles (not shown) during both ongoing activity ($b_{shuff}=0.67\pm 0.06$ vs. $b_{data}=0.60\pm 0.01$; mean $\pm$ SD, Mann-Whitney test, $p<0.001$, $20$ bootstraps), and evoked activity ($b_{shuff}=0.36\pm 0.07$ vs. $b_{data}=0.29\pm 0.01$; $p<0.001$). Especially during ongoing activity, this result was accompanied by a narrower distribution of pair-wise correlations in the shuffled datasets compared to the simultaneously recorded datasets (Fig. \ref{figthree}G), and is consistent with an inverse relationship between dimensionality and pair-wise correlations (see e.g. Eq. (\ref{eqnine})).

\subsection{Time course of dimensionality as a function of ensemble size}
\label{timecourse}
Unlike ongoing activity, the dependence of dimensionality on ensemble size (the slope of the linear regression of $d$ vs. $N$) was modulated during different epochs of the post-stimulus period (Fig. \ref{figthree}F, full lines; two-way ANOVA; main effect of time $F_{(4,495)}=3.80$, $p<0.005$; interaction time $\times$ condition: $F_{(4,495)}=4.76$, $p<0.001$). In particular, the dependence of $d$ on the ensemble size $N$ almost disappeared immediately after stimulus presentation in the simultaneously recorded, but not in the shuffled ensembles (trial-matched slope in the first evoked second: $b_{evoked}=0.07\pm 0.01$ vs $b_{shuff}=0.19\pm 0.07$) and converged to a stable value after approximately $1$ second (slope after the first second $b_{evoked}=0.38\pm 0.01$; compare with a stable average slope during ongoing activity of $b_{ongoing}=0.57\pm 0.01$, Fig. \ref{figthree}F).  
Note that the dimensionality is larger when the firing rate is computed in bins (as in Fig. \ref{figthree}F) rather than in HMM states (as in Fig. \ref{figthree}B-D, where the slopes are about half than in Fig. \ref{figthree}F). The reason is that firing rates and correlations are approximately constant during the same HMM state, whereas they may change when estimated in bins of fixed duration that include transitions among hidden states. These changes tend to dilute the correlations resulting in higher dimensionality as predicted e.g. by Eq. (\ref{eqnine}). A comparison of the pair-wise correlations of binned firing rates (Fig. \ref{figthree}G) vs. those of firing rates in HMM states (Fig. \ref{figthree}H) confirmed this hypothesis. Also, if the argument above is correct, one would expect a dependence of dimensionality on (fixed) bin duration. We computed the correlations and dimensionality of binned firing rates for various bin durations and found that $r$ increases and $d$ decreases for increasing bin durations (not shown). However, the slope of $d$ vs. $N$ is always larger in ongoing than in evoked activity regardless of bin size (ranging from $10$ ms to $5$ s; not shown). This confirms the generality of the results of Fig. \ref{figthree}B-D, which were obtained using firing rate vectors in hidden states. 
To summarize our main results so far, we found that dimensionality depends on ensemble size during both ongoing and evoked activity, and such dependence is significantly reduced in the post-stimulus period. This suggests that while state sequences during ongoing activity explore a large portion of the available firing rate space, the presentation of a stimulus initially collapses the state sequence along a more stereotyped and lower-dimensional response \cite{katz2001dynamic,jezzini2013processing}. During both ongoing and evoked activity, the dimensionality is also different than expected by chance in a set of independent neurons (shuffled datasets).

\begin{figure}[t]
\vspace*{-1.5cm}                                                           
\begin{center}
\hspace*{-0.8cm}                                                           
\includegraphics[width=1.1\textwidth]{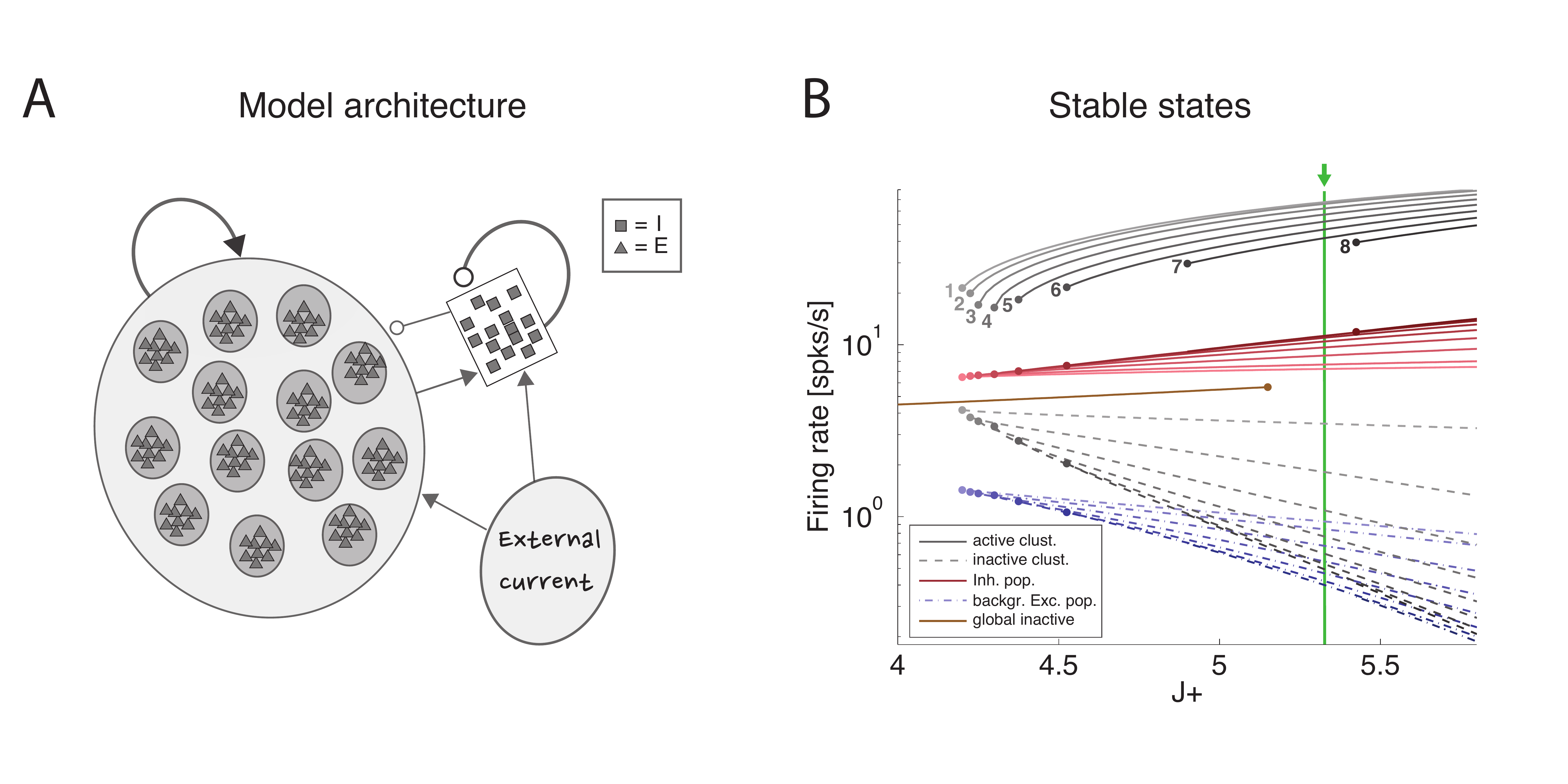}
\end{center}
\vspace*{-1cm}
\caption{{\it Recurrent network model}. A: Schematic recurrent network architecture. Triangles and squares represent excitatory and inhibitory LIF neurons respectively. Darker disks indicate excitatory clusters with potentiated intra-cluster synaptic weights. B: Mean field solution of the recurrent network. Firing rates of the stable states for each subpopulation are shown as function of the intra-cluster synaptic potentiation parameter $J_+$: firing rate activity in the active clusters (solid grey lines), firing rate in the inactive clusters (dashed grey lines), activity of the background excitatory population (dashed blue lines), activity of the inhibitory population (solid red lines). In each case, darker colors represent configurations with larger number of active clusters. Numbers denote the number of active clusters in each stable configuration. Configurations with $1$ to $8$ active clusters are stable in the limit of of infinite network size. A global configuration where all clusters are inactive (brown line) becomes unstable at the value $J_+=5.15$. The vertical green line represents the value of $J_+=5.3$ chosen for the simulations. X-axis: intra-cluster potentiation parameter $J_+$ in units of $J_{EE}$; Y-axis: Firing rate (spks/s).}
\label{figfour}
\end{figure}

\subsection{Clustered spiking network model of dimensionality}

\begin{figure}[t]
\begin{center}
\hspace*{-0.4cm}                                                           
\includegraphics[width=1.02\textwidth]{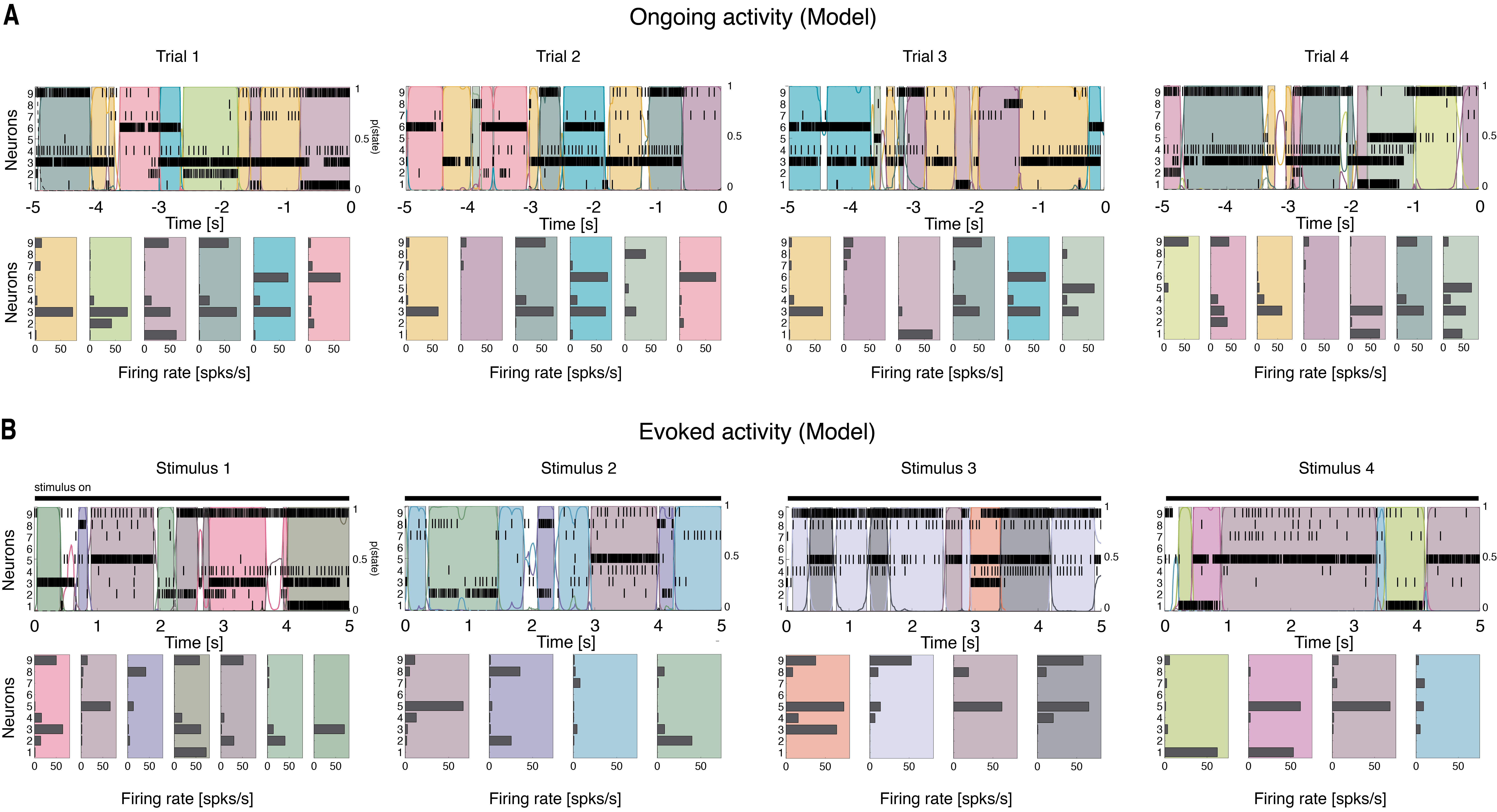}
\end{center}
\vspace*{-0.5cm}
\caption{{\it Ensemble activity in the recurrent network model is characterized by sequences of states}. Representative trials from one ensemble of nine simultaneously recorded neurons sampled from the recurrent network, segmented according to their ensemble states (notations as in Fig.~\ref{figone}). A: Ongoing activity. B: Ensemble activity evoked by four different stimuli, modeled as an increase in the external current to selected clusters (the black line on top of the raster plot represents the ``stimulus-on'' period; see Methods for details).}
\label{figfive}
\end{figure}

To gain a mechanistic understanding of the different dimensionality of ongoing and evoked activity we have analyzed a spiking network model with clustered connectivity which has been shown to capture many essential features of the data \cite{mazzucato2015dynamics}. In particular, the model reproduces the transitions among latent states in both ongoing and evoked activity. The network (see Section~ \ref{spikingmodel} for details) comprises $Q$ clusters of excitatory neurons characterized by stronger synaptic connections within each cluster and weaker connections between neurons in different clusters. All neurons receive recurrent input from a pool of inhibitory neurons that keeps the network in a balanced regime of excitation and inhibition in the absence of external stimulation (Fig.~\ref{figfour}A). In very large networks (technically, in networks with an infinite number of neurons), the stable configurations of the neural activity are characterized by a finite number of active clusters whose firing rates depend on the number clusters active at any given moment, as shown in Fig.~\ref{figfour}B (where $Q=30$). In a finite network, however, finite size effects ignite transitions among these configurations, inducing network states (firing rate vectors) on randomly chosen subsets of neurons that resemble the HMM states found in the data (Fig.~\ref{figfive}; see \cite{mazzucato2015dynamics} for details).

\begin{figure}[t]
\begin{center}
\vspace*{-0.8cm}                                                           
\hspace*{-2cm}                                                           
\includegraphics[width=1.24\textwidth]{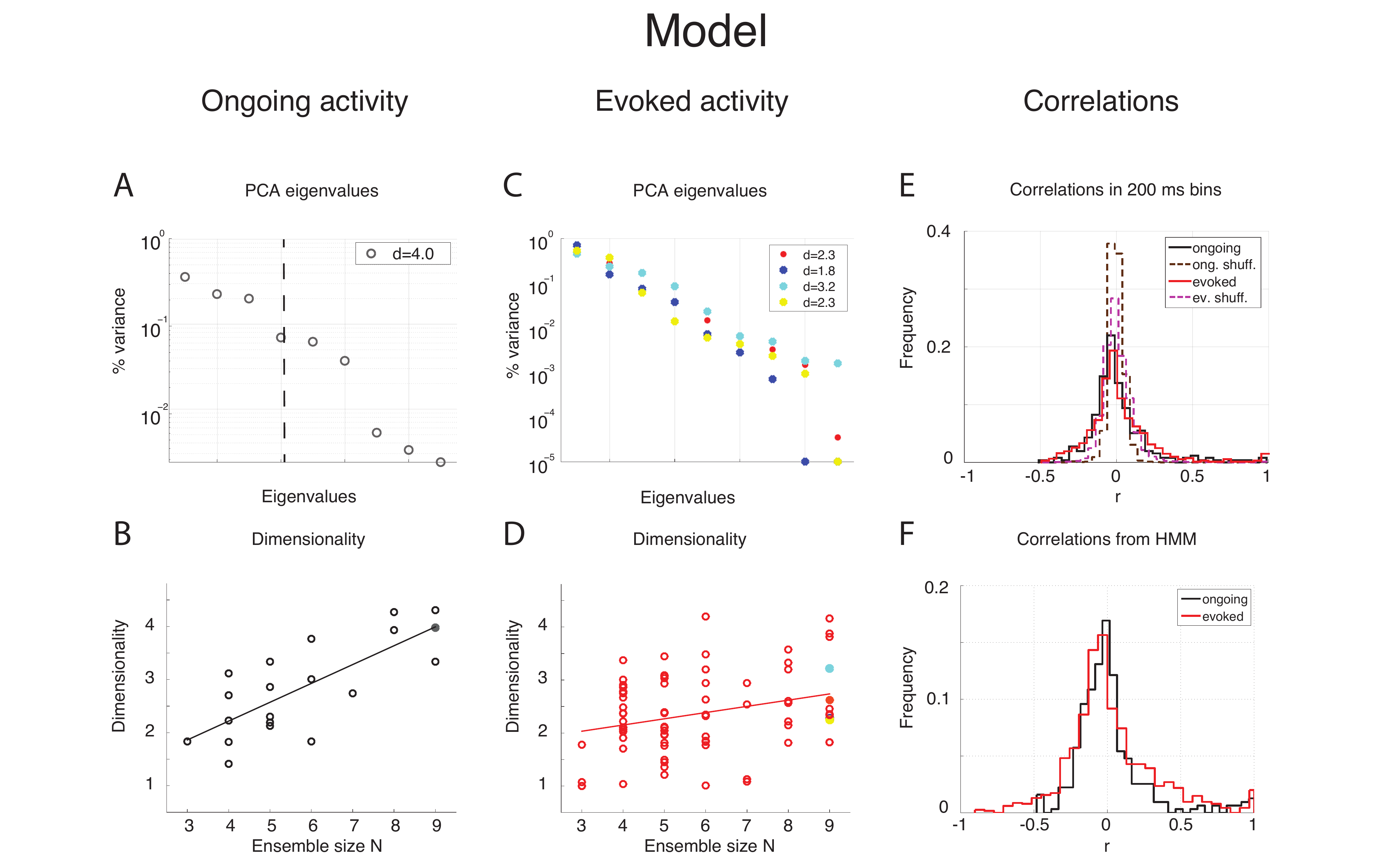}
\end{center}
\vspace*{-0.5cm}
\caption{{\it Dependence of dimensionality on ensemble size (model)}. A: Fraction of variance explained by each principal eigenvalue for an ensemble of 9 neurons during ongoing activity (corresponding to the filled dot in panel B) in the model network of Fig.~\ref{figfive} (notations as in Fig.~\ref{figthree}A). B: Dimensionality of neural activity across all ensembles in the model during ongoing activity  (linear regression fit, $d=b\cdot N+a$, $b=0.36\pm 0.07$, $a=0.80\pm 0.43$, $r=0.77$), estimated from HMM firing rate fits. X-axis: ensemble size; Y-axis: dimensionality. C: Fraction of variance explained by each principal eigenvalue for the ensemble in panel A during evoked activity. Principal eigenvalues for four stimuli are presented (corresponding to the color-coded dots in panel D). X-axis: eigenvalue number; Y-axis: percentage of variance explained by each eigenvalue. D: Dimensionality of neural activity across all ensembles in the model during evoked activity  (notations as in panel B, linear regression: $d=b\cdot N+a$, $b=0.12\pm 0.04$, $a=1.70\pm 0.26$, $r=0.29$). E: Distribution of pair-wise correlations in simultaneously recorded ensembles from the clustered network model (black and red histograms for ongoing and evoked activity, respectively) and in shuffled ensembles (brown and pink dashed histograms for ongoing and evoked activity, respectively) from $200$ ms bins. X-axis: correlation; Y-axis: frequency. F: Distribution of pair-wise correlations from HMM states during ongoing (black) and evoked activity (red) for all simultaneously recorded pairs of neurons. X-axis: correlation; Y-axis: frequency.}
\label{figsix}
\end{figure}

The dimensionality of the simulated sequences during ongoing and evoked activity was computed as done for the data, finding similar results. For the examples in Fig.~\ref{figfive}, we found $d_{ongoing}=4.0$ for ongoing activity (Fig.~\ref{figsix}A) between $d_{evoked}=2.2$ and $d_{evoked}=3.2$ across tastes during evoked activity (Fig.~\ref{figsix}C). Across all simulated sessions, we found an average $d_{ongoing}=2.9\pm0.9$ (mean$\pm$SD) for ongoing activity and $d_{evoked}=2.4\pm0.7$ for evoked activity. The model captured the essential properties of dimensionality observed in the data: the dimensionality did not differ across different tastes (one-way ANOVA, $p>0.2$) and depended on ensemble size during both ongoing (Fig.~\ref{figsix}B; slope $=0.36\pm0.07$, $r=0.77$, $p<10^{-4}$) and evoked periods (Fig.~\ref{figsix}D; slope $=0.12\pm0.04$, $r=0.29$, $p=0.01$). As for the data, the dependency on ensemble size was smaller for evoked compared to ongoing activity. We performed a trial-matched two-way ANOVA as done on the data and found, also in the model, a main effect of condition (ongoing vs. evoked: $F_{1,146}=22.1$, $p<10^{-5}$), a main effect of ensemble size ($F_{6,146}=14.1$, $p<10^{-11}$), and a significant interaction ($F_{6,146}=3.8$, $p=0.001$). These results were accompanied by patterns of correlations among the model neurons (Fig. \ref{figsix}E-F) very similar to those found in the data (Fig. \ref{figthree}G-H; see Section \ref{dimclusters} for statistics of correlation values). As in the data, narrower distributions of correlations were found for binned firing rates (Fig. \ref{figsix}E) compared to firing rates in hidden states (Fig. \ref{figsix}F; compare with Figs. \ref{figthree}G-H, respectively). Moreover, shuffling neurons across datasets reduced the correlations (Fig. \ref{figsix}E, dashed), resulting in a larger slope of $d$ vs. $N$ (not shown). Finally, $d$ during ongoing activity was always larger than during evoked activity also when computed on binned firing rates (not shown), as found in the data.
Since the model was not fine-tuned to find these results, the different dimensionalities of ongoing and evoked activity, and their associated patterns of pair-wise correlations, are likely the consequence of the organization in clusters and of the ensuing dynamics during ongoing and evoked activity.

\subsection{Scaling of dimensionality with ensemble size and pair-wise correlations}

The dependence of dimensionality on ensemble size observed in the data (Fig.~\ref{figthree}B) and in the model (Fig.~\ref{figsix}B) raises the question of whether or not the dimensionality would converge to an upper bound as one increases the number of simultaneously recorded neurons. In general, this question is important in a number of settings, related e.g. to coding in motor cortex \cite{ganguli2008one,gao2015simplicity}, performance in a discrimination task \cite{rigotti2013importance}, or coding of visual stimuli \cite{cadieu2013neural}. We can attack this question aided by the model of Fig. \ref{figfour}, where we can study the effect of large numbers of neurons, but also the impact on dimensionality of a clustered network architecture compared to a homogeneous one, at parity of correlations and ensemble size. 

\begin{figure}[ht]
\vspace*{-1.8cm}                                                           
\begin{center}
\hspace*{-0.8cm}                                                           
\includegraphics[width=1\textwidth]{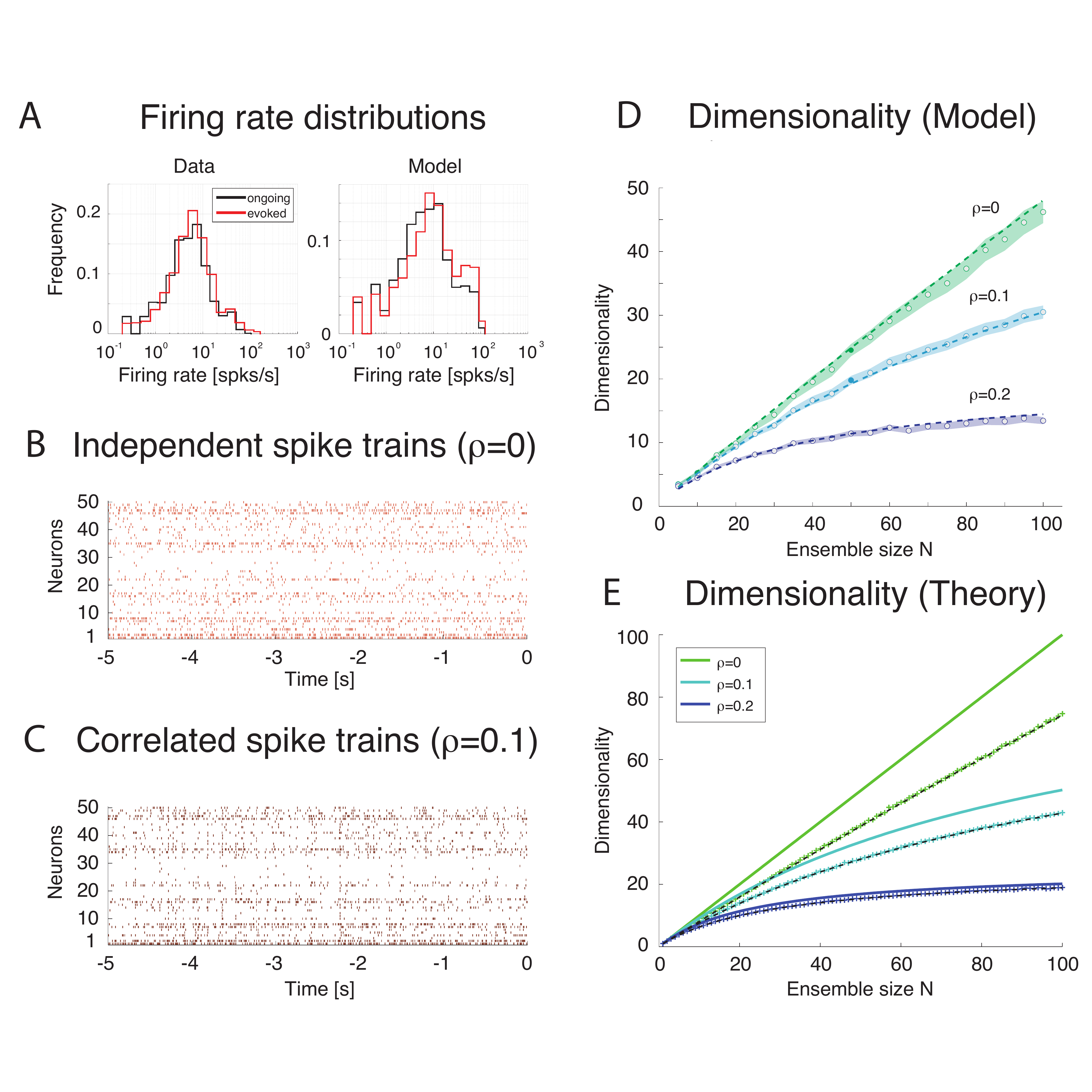}
\end{center}
\vspace*{-1cm}
\caption{{\it Dimensionality and correlation}. A: Empirical single neuron firing rate distributions in the data (left) and in the model (right), for ongoing (black) and evoked activity (red). The distributions are approximately lognormal. X-axis: Firing rate (spks/s); Y-axis: density. B: Example of independent Poisson spike trains with firing rates matched to the firing rates obtained in simulations of the spiking network model. C: Example of correlated Poisson spike trains with firing rates matched to the firing rates obtained in simulations of the spiking network model. Pair-wise correlations of  $\rho=0.1$ were used (see Methods). X-axis: time [s]; Y-axis: neuron index. D: Dimensionality as a function of ensemble size $N$ in an ensemble of Poisson spike trains with spike count correlations $\rho=0,0.1,0.2$ and firing rates matched to the model simulations of Fig. \ref{figsix}. Dashed lines represent the fit of Eq. (\ref{eqsixteen}) to the data (with $\delta\rho^2=\alpha\rho^2,\sigma^4=\delta\sigma^4=\beta$), with best-fit parameters (mean$\pm$s.e.m.) $\alpha=0.22\pm10^{-5}$, $\beta=340\pm 8$. Filled circles (from top to bottom): dimensionality of the data (raster plots) shown in panel B, C (shaded areas represent SD). X-axis: ensemble size; Y-axis, dimensionality. E: Theoretical prediction for the dependence of dimensionality on ensemble size N and spike count correlation $\rho$ for the case of uniform correlation, Eq. (\ref{eqeight}) (thick lines; green to cyan to blue shades represent increasing correlations).  ``$+$" are dimensionality estimates from $N_T=1,000$ trials for each $N$ (same $N_T$ as in panel D, each trial providing a firing rate value sampled from a log-normal distribution), in the case of log-normally distributed firing rate variances $\sigma_i^2$ with mean $\sigma^2=40$ (spk/s)$^2$ and standard deviation $0.5 \sigma^2$. Theoretical predictions from Eq. (\ref{eqsixteen}) match the estimated values in all cases (dashed black lines). X-axis: ensemble size N; Y-axis: dimensionality. }
\label{figseven}
\end{figure}

We consider first the case of a homogeneous network of neurons having no clusters and low pair-wise correlations, but having the same firing rates distributions (which were approximately log-normal, Fig. \ref{figseven}A) and the same mean pair-wise correlations as found in the data ($\rho \sim 0.01-0.2$). This would require solving a homogeneous recurrent network self-consistently for the desired firing rates and correlations. As a proxy for this scenario, we generated $20$ sessions of $40$ Poisson spike trains having exactly the desired properties (including the case of independent neurons for which $\rho =0$). Two examples with $\rho =0$ and $\rho =0.1$, respectively, are shown in Fig. \ref{figseven}B-C. Since in the asynchronous homogeneous network there are no transitions and hence no hidden states, the dimensionality was estimated based on the rate vectors in bins of $200$ ms duration (using bin widths of $50$ to $500$ ms did not change the results; see Methods for details).

We found that the dimensionality grows linearly with ensemble size in the absence of correlations, but is a concave function of $N$ in the presence of spike count correlations (circles in Fig.~\ref{figseven}D). Thus, as expected, the presence of correlations reduces the dimensionality. A simple theoretical calculation mimicking this scenario shows that d in this case converges slowly to an upper bound that depends on the inverse of the square of the pair-wise correlations. For example, in the case of uniform correlations ($\rho$) and equal variances of the spike counts, Eq. (\ref{eqeight}) of Methods,  $d(N,\rho)={1\over \rho^2+(1-\rho^2)/N}$, shows that $d=N$ in the absence of correlations, and $d\leq1/\rho^2$  for large networks in the presence of correlations. These properties remain approximately true in the case where the variances $\sigma_i^2$ of the spike counts are drawn from a distribution with mean $E[\sigma_i^2 ]=\sigma^2$ and variance $\textrm{Var}[\sigma_i^2 ]=\delta\sigma^4$. As Eq. (\ref{eqnine}) shows, in such a case dimensionality is reduced compared to the case of equal variances, for example $d\simeq {\sigma^4\over \sigma^4+\delta\sigma^4 } N<N$ for large $N$ when $\rho=0,\delta\rho=0$.

The analytical results are shown in Fig. \ref{figseven}E (full lines correspond to Eq. (\ref{eqeight})), together with their estimates (``+") based on $1,000$ data points (same number as trials in Fig. \ref{figseven}D; see Methods). The estimates are based on surrogate datasets with lognormal-distributed variances $\sigma_i^2$ to mimic the empirical distribution of variances found in GC (not shown).

\begin{figure}[ht]
\begin{center}
\vspace*{-1cm}                                                           
\hspace*{-0.7cm}                                                           
\includegraphics[width=1.05\textwidth]{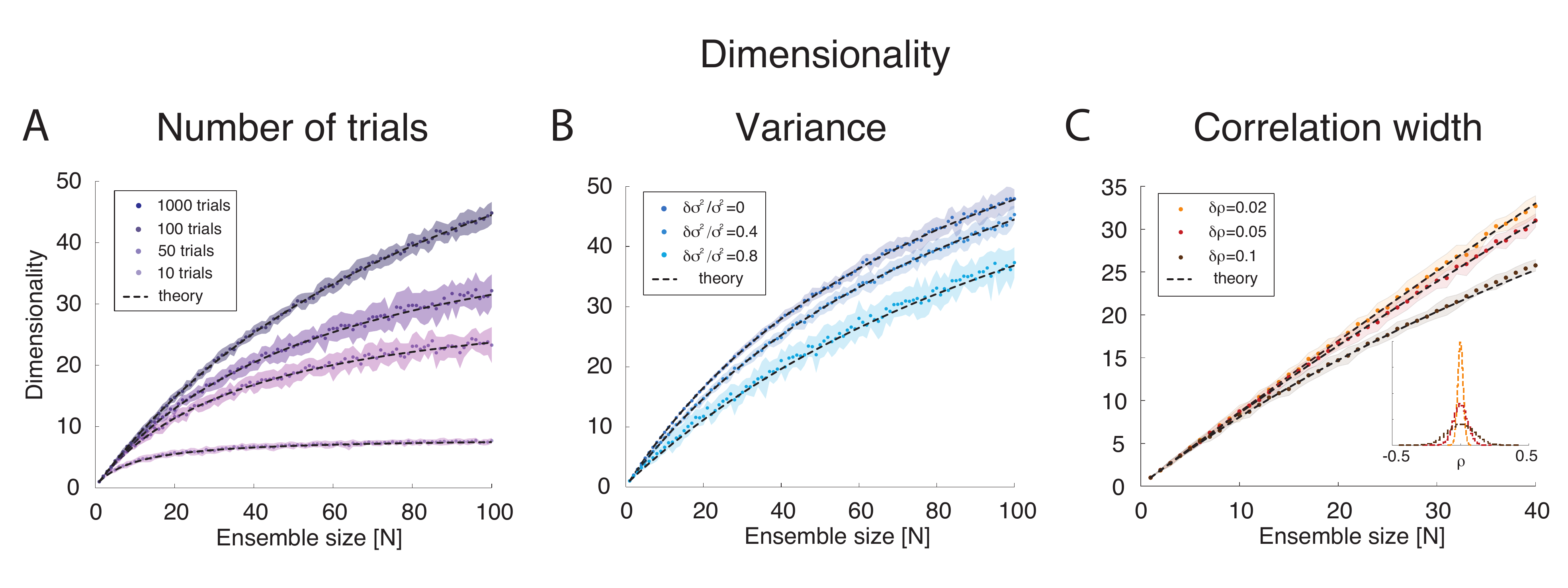}
\end{center}
\vspace*{-0.7cm}
\caption{{\it Dimensionality estimation}. A: Dependence of dimensionality on the number of trials for variable ensemble size $N$, for fixed correlations $\rho =0.1$ and firing rates variances $\sigma _i^2$ with mean $\sigma^2$ and standard deviation $\delta \sigma^2=0.4 \sigma ^2$. Dashed lines: theoretical prediction, Eq. (\ref{eqsixteen}); dots: mean values from simulations of $20$ surrogate datasets containing $10$ to $1000$ trials each (shaded areas:  SD), with darker shades representing increasing number of trials. X-axis: ensemble size; Y-axis, dimensionality. B: Dependence of dimensionality on the spread of the firing rates variances for fixed correlations $\rho =0.1$ and firing rate variance with mean $\sigma^2$. Dashed lines: theoretical prediction, Eq. (\ref{eqsixteen}); dots: mean values from simulations of $20$ surrogate datasets containing $1000$ trial each (shaded areas: SD), with lighter shades representing increasing values of $\delta \sigma^2/\sigma^2$). X-axis: ensemble size; Y-axis, dimensionality. C: Dependence of dimensionality on the width $\delta \rho =\sqrt{\textrm{Var}(\rho )}$ of pair-wise firing rate correlations (with zero mean), for firing rates variances $\sigma _i^2$ with mean $\sigma^2$ and standard deviation $\delta \sigma^2=0.4 \sigma^2$. Dashed lines: theoretical prediction, Eq. (\ref{eqsixteen}); dots: mean values from simulations of $20$ surrogate datasets containing $1000$ trials each (shaded areas: SD), with darker shades representing increasing values of $\delta \rho$. Inset: distribution of correlation coefficients used in the main figure. X-axis: ensemble size; Y-axis, dimensionality. In all panels, $\sigma^2=40$ (spk/s)$^2$.}
\label{figeight}
\end{figure}

\subsection{Estimation bias}

Comparison of Fig. \ref{figseven}D with \ref{figseven}E shows that the dimensionality of the homogeneous network is underestimated compared to the theoretical value given by Eq. (\ref{eqeight}). This is due to a finite number of trials and the presence of unequal variances with spread $\delta \sigma ^4$ (``+" in Fig. \ref{figseven}E). As Fig. \ref{figseven}E shows, taking this into account will reduce the dimensionality to values comparable to those of the homogeneous network of Fig. \ref{figseven}D. The dimensionality in that case is well predicted by Eq. (\ref{eqsixteen}) (broken lines in Fig. \ref{figseven}E). The same Eq. (\ref{eqsixteen}) was fitted successfully to the data in Fig. \ref{figseven}D (dashed) by tuning $2$ parameters to account for the unknown variance and correlation width of the firing rates (see Methods for details). 

Empirically, estimates of the dimensionality Eq. (\ref{eqtwo}) based on a finite number $N_T$ of trials tend to underestimate $d$ (Fig. \ref{eqthree}F). The approximate estimator Eq. (\ref{eqsixteen}) confirms that, for any ensemble size $N$, $d$ is a monotonically increasing function of the number of trials (Fig. \ref{figeight}A). Note that this holds for the mean value of the estimator (Eq. (\ref{eqsixteen})) over many datasets, not for single estimates, which could overestimate the true $d$ (not shown). Eq. (\ref{eqsixteen}) also provides an excellent description of dimensionality as a function of firing ratesÕ variance $\delta \sigma ^4$ (Fig. \ref{figeight}B) and pair-wise correlations width $\delta \rho^2$ (Fig. \ref{figeight}C). In particular, the mean and the variance of the pair-wise correlations have an interchangeable effect on $d$ (see Eq. (\ref{eqsixteen})); they both decrease the dimensionality and so does the firing rate variance $\delta \sigma ^4$ (Fig. \ref{figeight}B).

\subsection{Scaling of dimensionality in the presence of clusters}

We next compared the dimensionality of the homogeneous networkÕs activity to that predicted by the clustered network model of Fig.~\ref{figfour}. To allow comparison with the homogeneous network, dimensionality was computed based on the spike counts in $200$ ms bins rather than the HMMÕs firing rate vectors as in Fig.~\ref{figsix} (see Section~\ref{dimensionalitymeasure} for details). 

\begin{figure}[t]
\begin{center}
\vspace*{-1.2cm}                                                           
\hspace*{-0.2cm}                                                           
\includegraphics[width=1.02\textwidth]{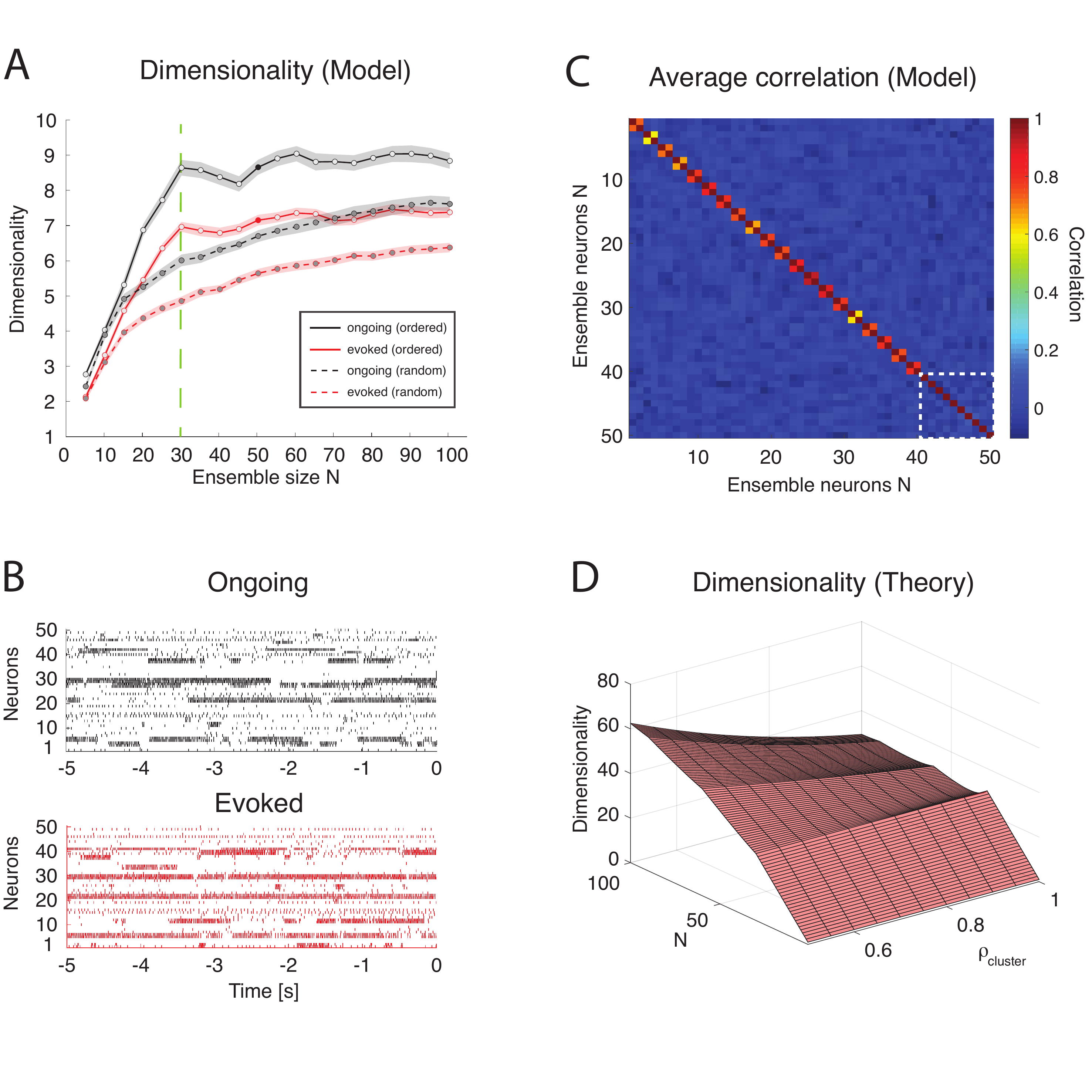}
\end{center}
\vspace*{-1.2cm}
\caption{{\it Dimensionality in a clustered network}. A: Trial-matched dimensionality as a function of ensemble size in the recurrent network model (ongoing and evoked activity in black and red, respectively, with shaded areas representing s.e.m.). Filled lines represent ordered sampling, where ensembles to the left of the green vertical line ($N=Q=30$) contain at most one neuron per cluster, while to the right of the line they contain one or more neurons from all clusters (filled circles indicate representative trials in panel B). Dashed lines represent random sampling of neurons, regardless of cluster membership. X-axis: ensemble size; Y-axis, dimensionality. B: Representative trial of an ensemble of $50$ neurons sampled from the recurrent network in Fig.~\ref{figfour} during ongoing activity (upper plot, in black) or evoked activity (lower plot, in red) for the case of Ôordered samplingÕ (full lines). Neurons are sorted according to their cluster membership (adjacent neuron pairs with similar activity belong to the same cluster, for neurons $\#1$ up to $\#40$; the last ten neurons are sampled from the remaining clusters). X-axis: time to stimulus presentation at $t=0$ (s); Y-axis: neuron index. C: Average correlation matrix for twenty ensembles of $N=50$ neurons from the clustered network model with $Q=30$. For the first $40$ neurons, adjacent pairs belong to the same cluster; the last $10$ neurons (delimited by a dashed white square) belong to the remaining clusters (neurons are ordered as in panel B). Thus, neurons $1, 3, 5,\ldots, 39$ ($20$ neurons) belong to the first $20$ clusters; neurons $2, 4, 6,\ldots, 40$ ($20$ neurons) belong also the first $20$ clusters; and neurons $41, 42, 43,\ldots, 50$ ($10$ neurons) belong to the remaining $10$ clusters. X-axis, Y-axis: neuron index. D: Plot of Eq. \ref{eqtwelve} giving $d$ vs. $N$ and $\rho$ (uniform within-cluster correlations) for the sampling procedure of panel B. X-axis: ensemble size $N$; Y-axis: dimensionality.}
\label{fignine}
\end{figure}

We found that the dependence of $d$ on $N$ in the clustered network depends on how the neurons are sampled. If the sampling is completely random, so that any neuron has the same probability of being added to the ensemble regardless of cluster membership, a concave dependence on $N$ will appear, much like the case of the homogeneous network (Fig.~\ref{fignine}A, dashed lines). However, if neurons are selected one from each cluster until all clusters have been sampled once, then one neuron from each cluster until all clusters have been sampled twice, and so on, until all the neurons in the network have been sampled, then the dependence of d on $N$ shows an abrupt transition when $N=Q$, i.e., when the number of sampled neurons reaches the number of clusters in the network (Fig.~\ref{fignine}A, full lines; see Fig 8B for raster plots with $Q=30$ and $N=50$). In the following, we refer to this sampling procedure as ``ordered sampling", as a reminder that neurons are selected randomly from each cluster, but the clusters are selected in serial order.  For $N\leq Q$, the dimensionality grows linearly with ensemble size in both ongoing (slope $0.24\pm0.01$, $r=0.79$, $p<10^{-10}$, black line) and evoked periods (slope $0.19\pm0.01$, $r=0.84$, $p<10^{-10}$; red line), and was larger during ongoing than evoked activity (trial-matched two-way ANOVA, main effect: $F_{1,948}=168$, $p<10^{-30}$; interaction: $F_{5,948}=4.1$, $p<0.001$). 
These results are in keeping with the empirical and model results based on the HMM analysis (Fig.~\ref{figthree} and \ref{figsix}). However, in the case of ordered sampling, the dependence of dimensionality on ensemble size tends to disappear for $N\geq Q$ both during ongoing (slope $0.010\pm0.003$, $r=0.1$, $p<0.001$) and evoked periods (slope $0.009\pm0.002$, $r=0.13$, $p<10^{-4}$; Fig.~\ref{fignine}A, full lines). The average dimensionality over the range $30\leq N\leq 100$ was significantly larger for ongoing, $d_{ongoing}=8.74\pm0.06$, than for evoked activity, $d_{evoked}=7.15\pm0.04$ (trial-matched two-way ANOVA, main effect: $F_{1,2212}=488$, $p<10^{-30}$), confirming that dimensionality is larger during ongoing than evoked activity also in this case. The difference in dimensionality between ongoing and evoked activity also holds in the case of random sampling on the entire range of $N$ values (Fig.~\ref{fignine}A, dashed lines), confirming the generality of this finding.

\subsection{Dimensionality is larger in the presence of clusters}
\label{dimclusters}
Intuitively, the dimensionality saturates at $N=Q$ in the clustered network because additional neurons will be highly correlated with already sampled ones.  For $N\leq Q$, each new neuronÕs activity adds an independent degree of freedom to the neural dynamics and thus increases its dimensionality. For $Q>N$, additional neurons are highly correlated with an existing neuron, adding little or no additional contribution to $d$. Indeed, compared to the low overall correlations found across all neuron pairs in the data (and used as {\it desiderata} for the homogeneous network), neurons belonging to the same model cluster had a much higher spike count correlation of $\rho=0.92$ $[0.56,0.96]$ (median and $[25, 75]$-percentile), while neurons belonging to different clusters had a much lower correlation of $\rho\simeq 0$ $[-0.10,0.06]$. A negligible median correlation was typical: for example, negligible was the overall median correlation regardless of cluster membership ($\rho \simeq 0\, [-0.109,0.083]$); and the empirical correlation both during ongoing ($[-0.047,0.051]$, with rare maximal values of $\rho \sim0.5$), and evoked activity ($[-0.085,0.113]$, with rare maximal values of $\rho\sim0.9$). While we note the qualitative agreement of model and empirical correlations, we emphasize that these numbers were obtained using $200$ ms bins and that they were quite sensitive to bin duration. In particular, the maximal correlations (regardless of sign) were substantially reduced for smaller bin durations (not shown). 
Plugging these values into a correlation matrix reflecting the clustered architecture and the sampling procedure used in Fig.~\ref{fignine}B, we obtained the matrix shown in Fig.~\ref{fignine}C, where pairwise correlations depend on whether or not the neurons belong to the same cluster (for the first $40$ neurons, adjacent pairs belong to the same cluster; the last $10$ neurons belong to the remaining clusters). It is natural to interpret such correlation matrix as the noisy observation of a block-diagonal matrix such that neurons in the same cluster have uniform correlation while neurons from different clusters are uncorrelated. For such a correlation matrix the dimensionality can be evaluated exactly (see Eq. (\ref{eqtwelve})). In the approximation where all neurons have the same variance, this reduces to Eq. (\ref{eqthirteen}), i.e.
\be\label{eqtwelveresults}
d=\left\{\begin{array}{cc}
N \ ,                                    &N\leq Q \\
{N\over 1+m\rho^2[1-(Q-p)/N]} \ ,& N>Q
\end{array}
\right.
\ee
where $N=mQ+p$. This formula is plotted in Fig.~\ref{fignine}D for relevant values of $\rho$ and $N$ and it explains the origin of the abrupt transition in dimensionality at $Q=N$. (The reasons for a dimensionality lower than $Q$ for $N\leq Q$ in the data Ð see Fig. \ref{fignine}A Ð are, also in this case, the finite number of data points ($250$) used for its estimation and the non-uniform distributions of firing rate variances and correlations). 

Note that the formula also predicts cusps in dimensionality (which become local maxima for large $\rho$) whenever the ensemble size is an exact multiple of the number of clusters. This is also visible in the simulated data of Fig.~\ref{fignine}A, where local maxima seem to appear at $N=30,60,90$ with $Q=30$ clusters. It is also worth mentioning that for low intra-cluster correlations the dependence on $N$ predicted by Eq.~ (\ref{eqthirteen})  becomes smoother and the cusps harder to detect (not shown), suggesting that the behavior of a clustered network with weak clusters tends to converge to the behavior of a homogeneous asynchronous network - therefore lacking sequences of hidden states. Thus, the complexity of the network dynamics is reflected in how its dimensionality scales with $N$, assuming that one may sample one neuron per cluster (i.e., via ``ordered sampling"). 
Even though it is not clear how to perform ordered sampling empirically (see Discussion), this result is nevertheless useful since it represents an upper bound also in the case of random sampling (see Fig.~\ref{fignine}A, dashed lines). Eq. (\ref{eqthirteen}) predicts that $d\leq Q/\rho^2$, with this value reached asymptotically for large $N$. In the case of random sampling, growth to this bound is even slower (Fig.~\ref{fignine}A). For comparison, in a homogeneous network $d\leq 1/\rho^2$ from Eq. ~ (\ref{eqeight}), a smaller bound by a factor of $Q$. Finally, homogeneous dimensionality is dominated by clustered dimensionality also in the more realistic case of non-uniform variances, where similar bounds are found in both cases (see Section~ \ref{dimensionalitymeasure} for details).

\section{Discussion}
\label{discussion}

In this paper we have investigated the dimensionality of the neural activity in the gustatory cortex of alert rats. Dimensionality was defined as a collective property of ensembles of simultaneously recorded neurons that reflects the effective space occupied by the ensemble activity during either ongoing or evoked activity. If one represents ensemble activity in terms of firing rate vectors, whose dimension is the number of ensemble neurons $N$, then the collection of rate vectors across trials takes the form of a set of points in the $N$-dimensional space of firing rates. Roughly, dimensionality is the minimal number of dimensions necessary to provide an accurate description of such set of points, which may be localized on a lower-dimensional subspace inside the whole firing rate space.

One of the main results of this paper is that the dimensionality of evoked activity is smaller than that of ongoing activity, i.e., stimulus presentation quenches dimensionality. More specifically, the dimensionality is linearly related to the ensemble size, with a significantly larger slope during ongoing activity compared to evoked activity (compare Fig.~\ref{figthree}B and ~\ref{figthree}D). We explained this phenomenon using a biologically plausible, mechanistic spiking network model based on recurrent connectivity with clustered architecture. The model was recently introduced in \cite{mazzucato2015dynamics} to account for the observed dynamics of ensembles of GC neurons as sequences of metastable states, where each state is defined as a vector of firing rates across simultaneously recorded neurons. The model captures the reduction in trial-to-trial variability and the multiple firing rates attained by single neurons across different states observed in GC upon stimulus presentation. Here, the same model was found to capture also the stimulus-induced reduction of dimensionality. While the set of active clusters during ongoing activity varies randomly, allowing the ensemble dynamics to explore a large portion of firing rate space, the evoked set of active clusters is limited mostly to the stimulus-selective clusters only (see \cite{mazzucato2015dynamics} for a detailed analysis). The dynamics of cluster activation in the model thus explains the more pronounced dependence of dimensionality on ensemble size found during ongoing compared to evoked activity.

We presented a simple theory of how dimensionality depends on the number of simultaneously recorded neurons $N$, their firing rate correlations, their variance, and the number and duration of recording trials. We found that dimensionality increases with $N$ and decreases with the amount of pair-wise correlations among the neurons (e.g., Fig. \ref{figeight}C). At parity of correlations, dimensionality is maximal when all neurons have the same firing rate variance, and it decreases as the distribution of count variances becomes more heterogeneous (e.g., Fig. \ref{figeight}B). The estimation of dimensionality based on a finite dataset is an increasing function of the number of trials (Fig. \ref{figeight}A). Finally, introducing clustered correlations in the theory, and sampling one neuron per cluster as in Fig. \ref{fignine}B, results in cusps at values of $N$ that are multiples of the number of clusters (Fig. \ref{fignine}D), in agreement with the predictions of the spiking network model (Fig. \ref{fignine}A, full lines). 

\subsection{Dimensionality scaling with ensemble size}

The increased dimensionality with sample size, especially during ongoing activity, was found empirically in datasets with $3$ to $9$ neurons per ensemble, but could be extrapolated for larger $N$ in a spiking network model with homogeneous or clustered architecture. In homogeneous networks with finite correlations the dimensionality is predicted to increase sub-linearly with $N$ (Eq. \ref{eqeight}), whereas in the clustered network it may exhibit cusps at multiple values of the number of clusters (Fig. \ref{fignine}A), and would saturate quickly to a value that depends on the ratio of the number of clusters $Q$ and the amount of pair-wise correlations, $d\leq Q/\rho^2$. Testing this prediction requires the ability to sample neurons one from each cluster, until all clusters are sampled, and seems beyond the current recording techniques. However, looking for natural groupings of neurons based on response similarities could uncover spatial segregation of clusters \cite{kiani2015natural} and could perhaps allow sampling neurons according to this procedure. Moreover, the model predicts a slower approach to a similar bound also in the case of random sampling. 

Dimensionality in a homogeneous network is instead bounded by $1/\rho^2$, and hence it is a factor $Q$ smaller than in the clustered network. Dimensionality is maximal in a population of independent neurons ($\rho=0$), where it grows linearly with $N$; however, neurons of recurrent networks have wide-ranging correlations (see e.g. Fig. \ref{figsix}E-F and its empirical counterpart, Fig. \ref{figthree}G-H). Since the presence of even low correlations can dramatically reduce the dimensionality (see e.g. Fig. \ref{figseven}D), the neural activity in a clustered architecture can reach much higher values at parity of correlations, representing an intermediate case between a homogeneous network and a population of independent neurons. 

Evidence for the presence of spatial clusters has been recently been reported in the prefrontal cortex based on correlations analyses \cite{kiani2015natural}. An alternative possibility is that neural clusters are not spatially but functionally arranged, and cluster memberships vary with time and task complexity \cite{rickert2009dynamic}. Can our model provide indirect tools to help uncover the presence of clusters? A closer look at Figs. 6E-F reveal a small peak at large correlations due to the contribution of highly correlated neurons belonging to the same cluster. This peak would be absent in a homogenous network and thus is the signature of a clustered architecture. However, such peak is populated by only small fraction ($1/Q$) of the total number of neuron pairs, which hinders its empirical detection (no peak at large correlations is clearly visible in our data, see Fig. \ref{figthree}G-H).  

\subsection{Dimensionality and trial-to-trial variability}

Cortical recordings from alert animals show that neurons produce irregular spike trains with variable spike counts across trials \cite{shadlen1994noise,fontanini2008behavioral,moreno2014poisson}. Despite many efforts, it remains a key issue to establish whether variability is detrimental \cite{gur1997response,white2012suppression} or useful \cite{mcdonnell2011benefits} for neural computation.

Trial-to-trial variability is reduced during preparatory activity \cite{churchland2006neural}, during the presentation of a stimulus \cite{churchland2010stimulus}, or when stimuli are expected \cite{samuelsen2012effects}, a phenomenon that would not occur in a population of independent or homogeneously connected neurons \cite{litwinkumar2012slow}. Recent work has shown that the stimulus-induced reduction of trial-to-trial variability can be due to spike-frequency adaptation in balanced networks \cite{farkhooi2013cellular} or to slow dynamic fluctuations generated in a recurrent spiking networks with clustered connectivity \cite{deco2012neural,litwinkumar2012slow,mazzucato2015dynamics}. In clustered network models, slow fluctuations in firing rates across neurons can ignite metastable sequences of neural activity, closely resembling metastable sequences observed experimentally  \cite{abeles1995cortical,seidemann1996simultaneously,jones2007natural,kemere2008detecting,durstewitz2010abrupt,poncealvarez2012dynamics,mazzucato2015dynamics}. The slow, metastable dynamics of cluster activation produces high variability in the spike count during ongoing activity. While cluster activations occur at random times during ongoing activity periods, stimulus presentation locks cluster activation at its onset, leading to a decrease in trial-to-trial variability.

Similarly, a stimulus-induced reduction of dimensionality is obtained in the same model. In this case, preferred cluster activation due to stimulus onset generates an increase in pair-wise correlations that reduce dimensionality. Note that the two properties (trial-to-trial variability and dimensionality) are conceptually distinct. An ensemble of Poisson spike trains can be highly correlated (hence have low dimensionality), yet the Fano Factor of each spike train will still be $1$ (hence high), independently of the correlations among neurons. In a recurrent network, however, dimensionality and trial-to-trial variability may become intertwined and exhibit similar properties, such as the stimulus-induced reduction observed in a model with clustered connectivity.  A deeper investigation of the link between dimensionality and trial-to-trial variability in recurrent networks is left for future studies. 

\subsection{Alternative definitions of dimensionality}

Following \cite{abbott2011interactions} we have defined dimensionality (Eq. (\ref{eqtwo})) as the dimension of an effective linear subspace of firing rate vectors containing the most variance of the neural activity. It differs from the typical dimensionality reduction based on PCA in that the latter retains only the number of eigenvectors explaining a predefined amount of variance (see e.g. \cite{broome2006encoding,geffen2009neural}), because Eq. (\ref{eqtwo}) includes contribution from all eigenvalues. Moreover, we have computed the firing rate correlations in bins of variable width that match the duration of the HMM states. Although our main results do not depend on bin size (see Section \ref{timecourse}), the actual value of dimensionality decreases with increasing bin duration. Thus, any choice of bin size (e.g., $200$ ms in Fig. \ref{figthree}F-G) remains somewhat arbitrary. A better method is to use a variable bin size as dictated by the HMM analysis, as done in Fig. \ref{figthree}B-D. This method also prevents diluting correlations among firing rates that would occur if one neuron were to change state inside the current bin, because during a hidden state the firing rates of the neurons are constant (by definition). Thus, this provides a principled adaptive procedure for selecting the bin size and eliminates the dependence of dimensionality on the bin width used for the analysis. 

Other definitions of neural dimensionality have been proposed in the literature, which aim at capturing different properties of the neural activity, typically during stimulus-evoked activity. A measure of dimensionality related to ours, and referred to as ``complexity," was introduced in \cite{cadieu2013neural}. According to their definition, population firing rate vectors from all evoked conditions were first decomposed along their kernel Principal Components \cite{montavon2011kernel}. A linear classifier was then trained on an increasing number of leading PCs in order to perform a discrimination task, where the number of PCs used was defined as the complexity of the representation. In general, the classification accuracy improves with increasing complexity, and it may saturate when all PCs containing relevant features are used - with the remaining PCs representing noise or information irrelevant to the task. Reaching high accuracy at low complexity implies good generalization performance, i.e., the ability to classify novel variations of a stimulus in the correct category. Neural representations in monkey inferotemporal cortex (IT) were found to require lower complexity than in area V4, confirming ITÕs premier role in classifying visual objects despite large variations in shape, orientation and background \cite{cadieu2013neural}. Complexity relies on a supervised algorithm and is an efficient tool to capture the generalization properties of evoked representations (see e.g. \cite{dicarlo2012does}) for its relevance to visual object recognition).

A second definition of dimensionality, sometimes referred to as ``shattering dimensionality" in the Machine Learning literature, was used to assess the discrimination properties of the neural representation \cite{rigotti2013importance}. Given a set of $p$ firing rate vectors, one can split them into two classes (e.g. white and black colorings) in $2^p$ different ways, and train a classifier to learn as many of those binary classification labels as possible. The shattering dimensionality is then defined as (the logarithm of) the largest number of binary classifications that can be implemented. This measure of dimensionality was found to drop significantly in monkey prefrontal cortex during error trials in a recall task and thus predicts the ability of the monkey to correctly perform the task \cite{rigotti2013importance}. 

A flexible and informative neural representation is one that achieves a large shattering dimensionality (good discrimination) while keeping a low complexity (good generalization). Note that both complexity and shattering dimensionality represent measures of classification performance in task-related paradigms, and their definition requires a set of evoked conditions to be classified via a supervised learning algorithm. While both definitions could be applied to neural activity in our stimulus-evoked data, their interpretation cannot be extended to periods of ongoing activity, as the latter is not associated to desired targets in a way that can be learned by a classification algorithm. Since our main aim was to compare the dimensionality of ongoing and evoked activity, the unsupervised approach of \cite{abbott2011interactions} and their notion of ``effective" dimensionality was better suited for our analysis. A related definition of dimensionality has been used by \cite{gao2015simplicity} to investigate neural representations of movements in motor cortex. 

Many measures of dimensionality used in the literature (including ours and some of those discussed above) are based on pair-wise correlations. However, neural activity is known to give rise also to higher-order correlations \cite{martignon2000neural}. Given that the extent and relevance of higher-order correlations is actively debated \cite{schneidman2006weak,staude2010cubic}, it would be useful to include them in measures of dimensionality. This is left for a future study.

\subsection{Ongoing activity and task complexity}

The relationship between ongoing and stimulus-evoked activity has been linked to the functional connectivity of local cortical circuits, and their mutual relationship has been the object of both theoretical and experimental investigations, often with contrasting conclusions (e.g., \cite{arieli1996dynamics,tsodyks1999linking,kenet2003spontaneously,luczak2009spontaneous,tkacik2010optimal,berkes2011spontaneous,mazzucato2015dynamics}. Here, we have focused on the dimensionality of ongoing and evoked activity and have shown that neural activity during ongoing periods occupies a space of larger dimensionality compared to evoked activity. Although based on a different measure of dimensionality, recent results on the relation between the dimensionality of evoked activity and task complexity suggest that evoked dimensionality is roughly equal to the number of task conditions \cite{rigotti2013importance}. It is natural to ask whether the dimensionality of ongoing activity provides an estimate of the complexity of the hardest task that can be supported by the neural activity. Moreover, based on the clustered network model, the presence of clusters imposes an upper value $d\leq Q/\rho^2$ during ongoing activity, suggesting that a discrimination task with up to $\sim Q$ different conditions may be supported. The experience of taste consumption is by itself multidimensional, including both chemo- and oro-sensory aspects (i.e., taste identity \cite{jezzini2013processing} and concentration \cite{sadacca2012sodium}, texture, temperature \cite{yamamoto1981cortical,yamamoto1988sensory}, taste and odor mixtures \cite{maier2013neural}) and psychological aspects (hedonic value \cite{katz2001dynamic,grossman2008learning,jezzini2013processing}), anticipation \cite{samuelsen2012effects}, novelty \cite{bermudez2014forgotten} and satiety effects \cite{de2006neural}. It is tempting to speculate that neural activity during ongoing periods explores these different dimensions, while evoked activity is confined to the features of the particular taste being delivered in a specific context. Establishing a precise experimental and theoretical link between the number of clusters and task complexity is an important question left for future studies.

\bigskip
\noindent{\bf\large Conflict of Interest Statement}
\bigskip 

\noindent The authors declare that the research was conducted in the absence of any commercial or financial relationships that could be construed as a potential conflict of interest.

\section*{Acknowledgments}

This work was supported by a National Institute of Deafness and Other Communication Disorders Grant K25-DC013557 (LM), by the Swartz Foundation Award 66438 (LM), by National Institute of Deafness and Other Communication Disorders Grant R01-DC010389 (AF), by a Klingenstein Foundation Fellowship (AF), and by a National Science Foundation Grant IIS1161852 (GLC). We thank Dr. Stefano Fusi and Memming Park for useful discussions and David Ecker at the Research Technologies DoIT of Stony Brook University for access to its computational resources.

\bibliography{bib}

\providecommand{\href}[2]{#2}\begingroup\raggedright\begin{thebibliography}{10}

\bibitem{abeles1995cortical}
M.~Abeles, H.~Bergman, I.~Gat, I.~Meilijson, E.~Seidemann, N.~Tishby, and
  E.~Vaadia, {\it Cortical activity flips among quasi-stationary states},  {\em
  Proc Natl Acad Sci USA} {\bf 92} (1995) 8616--8620.

\bibitem{seidemann1996simultaneously}
E.~Seidemann, I.~Meilijson, M.~Abeles, H.~Bergman, and E.~Vaadia, {\it
  Simultaneously recorded single units in the frontal cortex go through
  sequences of discrete and stable states in monkeys performing a delayed
  localization task},  {\em J Neurosci} {\bf 16} (1996), no.~2 752--68.

\bibitem{durstewitz2010abrupt}
D.~Durstewitz, N.~M. Vittoz, S.~B. Floresco, and J.~K. Seamans, {\it Abrupt
  transitions between prefrontal neural ensemble states accompany behavioral
  transitions during rule learning},  {\em Neuron} {\bf 66} (2010), no.~3
  438--48.

\bibitem{jones2007natural}
L.~M. Jones, A.~Fontanini, B.~F. Sadacca, P.~Miller, and D.~B. Katz, {\it
  Natural stimuli evoke dynamic sequences of states in sensory cortical
  ensembles},  {\em Proc Natl Acad Sci U S A} {\bf 104} (2007), no.~47
  18772--7.

\bibitem{mazzucato2015dynamics}
L.~Mazzucato, A.~Fontanini, and G.~La~Camera, {\it Dynamics of multistable
  states during ongoing and evoked cortical activity},  {\em The Journal of
  Neuroscience} {\bf 35} (2015), no.~21 8214--8231.

\bibitem{kemere2008detecting}
C.~Kemere, G.~Santhanam, B.~M. Yu, A.~Afshar, S.~I. Ryu, T.~H. Meng, and K.~V.
  Shenoy, {\it Detecting neural-state transitions using hidden markov models
  for motor cortical prostheses},  {\em J Neurophysiol} {\bf 100} (2008), no.~4
  2441--52.

\bibitem{poncealvarez2012dynamics}
A.~Ponce-Alvarez, V.~Nacher, R.~Luna, A.~Riehle, and R.~Romo, {\it Dynamics of
  cortical neuronal ensembles transit from decision making to storage for later
  report},  {\em J Neurosci} {\bf 32} (2012), no.~35 11956--69.

\bibitem{ganguli2008one}
S.~Ganguli, J.~W. Bisley, J.~D. Roitman, M.~N. Shadlen, M.~E. Goldberg, and
  K.~D. Miller, {\it One-dimensional dynamics of attention and decision making
  in lip},  {\em Neuron} {\bf 58} (2008), no.~1 15--25.

\bibitem{churchland2010cortical}
M.~M. Churchland, J.~P. Cunningham, M.~T. Kaufman, S.~I. Ryu, and K.~V. Shenoy,
  {\it Cortical preparatory activity: representation of movement or first cog
  in a dynamical machine?},  {\em Neuron} {\bf 68} (2010), no.~3 387--400.

\bibitem{abbott2011interactions}
L.~F. Abbott, K.~Rajan, and H.~Sompolinsky, {\em Interactions between Intrinsic
  and Stimulus-Evoked Activity in Recurrent Neural Networks}, ch.~4.
\newblock Oxford University Press, 2011.

\bibitem{ganguli2012compressed}
S.~Ganguli and H.~Sompolinsky, {\it Compressed sensing, sparsity, and
  dimensionality in neuronal information processing and data analysis},  {\em
  Annual review of neuroscience} {\bf 35} (2012) 485--508.

\bibitem{cadieu2013neural}
C.~F. Cadieu, H.~Hong, D.~Yamins, N.~Pinto, N.~J. Majaj, and J.~J. DiCarlo,
  {\it The neural representation benchmark and its evaluation on brain and
  machine},  {\em arXiv preprint arXiv:1301.3530} (2013).

\bibitem{rigotti2013importance}
M.~Rigotti, O.~Barak, M.~R. Warden, X.~J. Wang, N.~D. Daw, E.~K. Miller, and
  S.~Fusi, {\it The importance of mixed selectivity in complex cognitive
  tasks},  {\em Nature} {\bf 497} (2013), no.~7451 585--90.

\bibitem{gao2015simplicity}
P.~Gao and S.~Ganguli, {\it On simplicity and complexity in the brave new world
  of large-scale neuroscience},  {\em Current opinion in neurobiology} {\bf 32}
  (2015) 148--155.

\bibitem{cohen2009attention}
M.~R. Cohen and J.~H. Maunsell, {\it Attention improves performance primarily
  by reducing interneuronal correlations},  {\em Nature neuroscience} {\bf 12}
  (2009), no.~12 1594--1600.

\bibitem{nienborg2012decision}
H.~Nienborg, M.~R.~Cohen, and B.~G. Cumming, {\it Decision-related activity in
  sensory neurons: correlations among neurons and with behavior},  {\em Annual
  review of neuroscience} {\bf 35} (2012) 463--483.

\bibitem{samuelsen2012effects}
C.~L. Samuelsen, M.~P. Gardner, and A.~Fontanini, {\it Effects of cue-triggered
  expectation on cortical processing of taste},  {\em Neuron} {\bf 74} (2012),
  no.~2 410--422.

\bibitem{fontanini2006state}
A.~Fontanini and D.~B. Katz, {\it State-dependent modulation of time-varying
  gustatory responses},  {\em J Neurophysiol} {\bf 96} (2006), no.~6 3183--93.

\bibitem{fontanini20057}
A.~Fontanini and D.~B. Katz, {\it 7 to 12 hz activity in rat gustatory cortex
  reflects disengagement from a fluid self-administration task},  {\em J
  Neurophysiol} {\bf 93} (2005), no.~5 2832--40.

\bibitem{phillips1970rapid}
M.~Phillips and R.~Norgren, {\it A rapid method for permanent implantation of
  an intraoral fistula in rats},  {\em Behavioral Research, Methods, and
  Instrumentation} (1970), no.~2 124.

\bibitem{fontanini2008behavioral}
A.~Fontanini and D.~B. Katz, {\it Behavioral states, network states, and
  sensory response variability},  {\em J Neurophysiol} {\bf 100} (2008), no.~3
  1160--8.

\bibitem{katz2001dynamic}
D.~B. Katz, S.~A. Simon, and M.~A. Nicolelis, {\it Dynamic and multimodal
  responses of gustatory cortical neurons in awake rats},  {\em J Neurosci}
  {\bf 21} (2001), no.~12 4478--89.

\bibitem{horst2013reward}
N.~K. Horst and M.~Laubach, {\it Reward-related activity in the medial
  prefrontal cortex is driven by consumption},  {\em Front Neurosci} {\bf 7}
  (2013) 56.

\bibitem{jezzini2013processing}
A.~Jezzini, L.~Mazzucato, G.~La~Camera, and A.~Fontanini, {\it Processing of
  hedonic and chemosensory features of taste in medial prefrontal and insular
  networks},  {\em The Journal of Neuroscience} {\bf 33} (2013), no.~48
  18966--18978.

\bibitem{escola2011hidden}
S.~Escola, A.~Fontanini, D.~Katz, and L.~Paninski, {\it Hidden markov models
  for the stimulus-response relationships of multistate neural systems},  {\em
  Neural Comput} {\bf 23} (2011), no.~5 1071--132.

\bibitem{rabiner1989tutorial}
L.~R. Rabiner, {\it A tutorial on hidden markov models and selected
  applications in speech recognition},  {\em Proceedings of the IEEE} {\bf 77}
  (1989) 257--286.

\bibitem{miller2010stochastic}
P.~Miller and D.~B. Katz, {\it Stochastic transitions between neural states in
  taste processing and decision-making},  {\em J Neurosci} {\bf 30} (2010),
  no.~7 2559--70.

\bibitem{renart2010asynchronous}
A.~Renart, J.~de~la Rocha, P.~Bartho, L.~Hollender, N.~Parga, A.~Reyes, and
  K.~D. Harris, {\it The asynchronous state in cortical circuits},  {\em
  Science} {\bf 327} (2010), no.~5965 587--90.

\bibitem{chapin1999principal}
J.~K. Chapin and M.~A. Nicolelis, {\it Principal component analysis of neuronal
  ensemble activity reveals multidimensional somatosensory representations},
  {\em Journal of neuroscience methods} {\bf 94} (1999), no.~1 121--140.

\bibitem{mardia1979multivariate}
K.~V. Mardia, J.~T. Kent, and J.~M. Bibby.
\newblock Academic Presss, 1979.

\bibitem{holland1977robust}
P.~W. Holland and R.~E. Welsch, {\it Robust regression using iteratively
  reweighted least-squares},  {\em Communications in Statistics-theory and
  Methods} {\bf 6} (1977), no.~9 813--827.

\bibitem{macke2009generating}
J.~H. Macke, P.~Berens, A.~S. Ecker, A.~S. Tolias, and M.~Bethge, {\it
  Generating spike trains with specified correlation coefficients},  {\em
  Neural Computation} {\bf 21} (2009), no.~2 397--423.

\bibitem{amit1997model}
D.~J. Amit and N.~Brunel, {\it Model of global spontaneous activity and local
  structured activity during delay periods in the cerebral cortex},  {\em Cereb
  Cortex} {\bf 7} (1997), no.~3 237--52.

\bibitem{brunel1999fast}
N.~Brunel and V.~Hakim, {\it Fast global oscillations in networks of
  integrate-and-fire neurons with low firing rates},  {\em Neural Comput} {\bf
  11} (1999), no.~7 1621--71.

\bibitem{fusi1999collective}
S.~Fusi and M.~Mattia, {\it Collective behavior of networks with linear (vlsi)
  integrate-and-fire neurons},  {\em Neural Comput} {\bf 11} (1999), no.~3
  633--52.

\bibitem{curti2004mean}
E.~Curti, G.~Mongillo, G.~La~Camera, and D.~J. Amit, {\it Mean field and
  capacity in realistic networks of spiking neurons storing sparsely coded
  random memories},  {\em Neural computation} {\bf 16} (2004), no.~12
  2597--2637.

\bibitem{tuckwell1988introduction}
H.~Tuckwell, {\em Introduction to theoretical neurobiology}, vol.~2.
\newblock Cambridge University Press, 1988.

\bibitem{lansky1999stochastic}
P.~Lansky and S.~Sato, {\it The stochastic diffusion models of nerve membrane
  depolarization and interspike interval generation},  {\em J Peripher Nerv
  Syst} {\bf 4} (1999), no.~1 27--42.

\bibitem{richardson2004effects}
M.~J. Richardson, {\it The effects of synaptic conductance on the voltage
  distribution and firing rate of spiking neurons.},  {\em Phys Rev E Stat
  Nonlin Soft Matter Phys} {\bf 69} (2004) 051,918.

\bibitem{van1996chaos}
C.~van Vreeswijk and H.~Sompolinsky, {\it Chaos in neuronal networks with
  balanced excitatory and inhibitory activity},  {\em Science} {\bf 274}
  (1996), no.~5293 1724--1726.

\bibitem{vreeswijk1998chaotic}
C.~v. Vreeswijk and H.~Sompolinsky, {\it Chaotic balanced state in a model of
  cortical circuits},  {\em Neural computation} {\bf 10} (1998), no.~6
  1321--1371.

\bibitem{Brunel1998}
N.~Brunel and S.~Sergi, {\it Firing frequency of leaky intergrate-and-fire
  neurons with synaptic current dynamics},  {\em J Theor Biol} {\bf 195}
  (1998), no.~1 87--95.

\bibitem{Fourcaud2002}
N.~Fourcaud and N.~Brunel, {\it Dynamics of the firing probability of noisy
  integrate-and-fire neurons},  {\em Neural Comput} {\bf 14} (2002), no.~9
  2057--110.

\bibitem{Rauch2003}
A.~Rauch, G.~La~Camera, H.~R. Luscher, W.~Senn, and S.~Fusi, {\it Neocortical
  pyramidal cells respond as integrate-and-fire neurons to in vivo-like input
  currents},  {\em J Neurophysiol} {\bf 90} (2003), no.~3 1598--612.

\bibitem{LaCamera2006}
G.~La~Camera, A.~Rauch, D.~Thurbon, H.~R. Luscher, W.~Senn, and S.~Fusi, {\it
  Multiple time scales of temporal response in pyramidal and fast spiking
  cortical neurons},  {\em J Neurophysiol} {\bf 96} (2006), no.~6 3448--64.

\bibitem{LaCamera2008}
G.~La~Camera, M.~Giugliano, W.~Senn, and S.~Fusi, {\it The response of cortical
  neurons to in vivo-like input current: theory and experiment : I. noisy
  inputs with stationary statistics},  {\em Biol Cybern} {\bf 99} (2008),
  no.~4-5 279--301.

\bibitem{giugliano2004single}
M.~Giugliano, P.~Darbon, M.~Arsiero, H.-R. L{\"u}scher, and J.~Streit, {\it
  Single-neuron discharge properties and network activity in dissociated
  cultures of neocortex},  {\em Journal of neurophysiology} {\bf 92} (2004),
  no.~2 977--996.

\bibitem{Press2007}
W.~Press, S.~Teukolsky, W.~Vetterling, and B.~Flannery, {\it Numerical recipes
  the art of scientific computing},  2007.

\bibitem{Mascaro1999}
M.~Mascaro and D.~J. Amit, {\it Effective neural response function for
  collective population states},  {\em Network} {\bf 10} (1999), no.~4 351--73.

\bibitem{LaCamera2004}
G.~La~Camera, A.~Rauch, H.~R. Luscher, W.~Senn, and S.~Fusi, {\it Minimal
  models of adapted neuronal response to in vivo-like input currents},  {\em
  Neural Comput} {\bf 16} (2004), no.~10 2101--24.

\bibitem{Renart2004}
A.~Renart, N.~Brunel, and X.-J. Wang, {\em Mean-field theory of recurrent
  cortical networks: from irregularly spiking neurons to working memory},
  pp.~431--490.
\newblock Boca Raton, FL: CRC, 2004.

\bibitem{Giugliano2008}
M.~Giugliano, G.~La~Camera, S.~Fusi, and W.~Senn, {\it The response of cortical
  neurons to in vivo-like input current: theory and experiment: Ii.
  time-varying and spatially distributed inputs},  {\em Biol Cybern} {\bf 99}
  (2008), no.~4-5 303--18.

\bibitem{kiani2015natural}
R.~Kiani, C.~J. Cueva, J.~B. Reppas, D.~Peixoto, S.~I. Ryu, and W.~T. Newsome,
  {\it Natural grouping of neural responses reveals spatially segregated
  clusters in prearcuate cortex},  {\em Neuron} {\bf 85} (2015), no.~6
  1359--1373.

\bibitem{rickert2009dynamic}
J.~Rickert, A.~Riehle, A.~Aertsen, S.~Rotter, and M.~P. Nawrot, {\it Dynamic
  encoding of movement direction in motor cortical neurons},  {\em The Journal
  of Neuroscience} {\bf 29} (2009), no.~44 13870--13882.

\bibitem{shadlen1994noise}
M.~N. Shadlen and W.~T. Newsome, {\it Noise, neural codes and cortical
  organization},  {\em Curr Opin Neurobiol} {\bf 4} (1994), no.~4 569--79.

\bibitem{moreno2014poisson}
R.~Moreno-Bote, {\it Poisson-like spiking in circuits with probabilistic
  synapses.},  {\em PLoS computational biology} {\bf 10} (2014), no.~7
  e1003522--e1003522.

\bibitem{gur1997response}
M.~Gur, A.~Beylin, and D.~M. Snodderly, {\it Response variability of neurons in
  primary visual cortex (v1) of alert monkeys},  {\em The Journal of
  neuroscience} {\bf 17} (1997), no.~8 2914--2920.

\bibitem{white2012suppression}
B.~White, L.~F. Abbott, and J.~Fiser, {\it Suppression of cortical neural
  variability is stimulus-and state-dependent},  {\em Journal of
  neurophysiology} {\bf 108} (2012), no.~9 2383--2392.

\bibitem{mcdonnell2011benefits}
M.~D. McDonnell and L.~M. Ward, {\it The benefits of noise in neural systems:
  bridging theory and experiment},  {\em Nature Reviews Neuroscience} {\bf 12}
  (2011), no.~7 415--426.

\bibitem{churchland2006neural}
M.~M. Churchland, M.~Y. Byron, S.~I. Ryu, G.~Santhanam, and K.~V. Shenoy, {\it
  Neural variability in premotor cortex provides a signature of motor
  preparation},  {\em The Journal of neuroscience} {\bf 26} (2006), no.~14
  3697--3712.

\bibitem{churchland2010stimulus}
M.~M. Churchland, B.~M. Yu, J.~P. Cunningham, L.~P. Sugrue, M.~R. Cohen, G.~S.
  Corrado, W.~T. Newsome, A.~M. Clark, P.~Hosseini, B.~B. Scott, D.~C. Bradley,
  M.~A. Smith, A.~Kohn, J.~A. Movshon, K.~M. Armstrong, T.~Moore, S.~W. Chang,
  L.~H. Snyder, S.~G. Lisberger, N.~J. Priebe, I.~M. Finn, D.~Ferster, S.~I.
  Ryu, G.~Santhanam, M.~Sahani, and K.~V. Shenoy, {\it Stimulus onset quenches
  neural variability: a widespread cortical phenomenon},  {\em Nat Neurosci}
  {\bf 13} (2010), no.~3 369--78.

\bibitem{litwinkumar2012slow}
A.~Litwin-Kumar and B.~Doiron, {\it Slow dynamics and high variability in
  balanced cortical networks with clustered connections},  {\em Nat Neurosci}
  {\bf 15} (2012), no.~11 1498--505.

\bibitem{farkhooi2013cellular}
F.~Farkhooi, A.~Froese, E.~Muller, R.~Menzel, and M.~P. Nawrot, {\it Cellular
  adaptation facilitates sparse and reliable coding in sensory pathways},  {\em
  PLoS Comput Biol} {\bf 9} (2013), no.~10 e1003251.

\bibitem{deco2012neural}
G.~Deco and E.~Hugues, {\it Neural network mechanisms underlying stimulus
  driven variability reduction},  {\em PLoS Comput Biol} {\bf 8} (2012), no.~3
  e1002395.

\bibitem{broome2006encoding}
B.~M. Broome, V.~Jayaraman, and G.~Laurent, {\it Encoding and decoding of
  overlapping odor sequences},  {\em Neuron} {\bf 51} (2006), no.~4 467--482.

\bibitem{geffen2009neural}
M.~N. Geffen, B.~M. Broome, G.~Laurent, and M.~Meister, {\it Neural encoding of
  rapidly fluctuating odors},  {\em Neuron} {\bf 61} (2009), no.~4 570--586.

\bibitem{montavon2011kernel}
G.~Montavon, M.~L. Braun, and K.-R. M{\"u}ller, {\it Kernel analysis of deep
  networks},  {\em The Journal of Machine Learning Research} {\bf 12} (2011)
  2563--2581.

\bibitem{dicarlo2012does}
J.~J. DiCarlo, D.~Zoccolan, and N.~C. Rust, {\it How does the brain solve
  visual object recognition?},  {\em Neuron} {\bf 73} (2012), no.~3 415--434.

\bibitem{martignon2000neural}
L.~Martignon, G.~Deco, K.~Laskey, M.~Diamond, W.~Freiwald, and E.~Vaadia, {\it
  Neural coding: higher-order temporal patterns in the neurostatistics of cell
  assemblies},  {\em Neural Computation} {\bf 12} (2000), no.~11 2621--2653.

\bibitem{schneidman2006weak}
E.~Schneidman, M.~J. Berry, R.~Segev, and W.~Bialek, {\it Weak pairwise
  correlations imply strongly correlated network states in a neural
  population},  {\em Nature} {\bf 440} (2006), no.~7087 1007--1012.

\bibitem{staude2010cubic}
B.~Staude, S.~Rotter, and S.~Gr{\"u}n, {\it Cubic: cumulant based inference of
  higher-order correlations in massively parallel spike trains},  {\em Journal
  of Computational Neuroscience} {\bf 29} (2010), no.~1-2 327--350.

\bibitem{arieli1996dynamics}
A.~Arieli, A.~Sterkin, A.~Grinvald, and A.~Aertsen, {\it Dynamics of ongoing
  activity: explanation of the large variability in evoked cortical responses},
   {\em Science} {\bf 273} (1996), no.~5283 1868--1871.

\bibitem{tsodyks1999linking}
M.~Tsodyks, T.~Kenet, A.~Grinvald, and A.~Arieli, {\it Linking spontaneous
  activity of single cortical neurons and the underlying functional
  architecture},  {\em Science} {\bf 286} (1999), no.~5446 1943--1946.

\bibitem{kenet2003spontaneously}
T.~Kenet, D.~Bibitchkov, M.~Tsodyks, A.~Grinvald, and A.~Arieli, {\it
  Spontaneously emerging cortical representations of visual attributes},  {\em
  Nature} {\bf 425} (2003), no.~6961 954--6.

\bibitem{luczak2009spontaneous}
A.~Luczak, P.~Bartho, and K.~D. Harris, {\it Spontaneous events outline the
  realm of possible sensory responses in neocortical populations},  {\em
  Neuron} {\bf 62} (2009), no.~3 413--25.

\bibitem{tkacik2010optimal}
G.~Tkacik, J.~S. Prentice, V.~Balasubramanian, and E.~Schneidman, {\it Optimal
  population coding by noisy spiking neurons},  {\em Proc Natl Acad Sci U S A}
  {\bf 107} (2010), no.~32 14419--24.

\bibitem{berkes2011spontaneous}
P.~Berkes, G.~Orban, M.~Lengyel, and J.~Fiser, {\it Spontaneous cortical
  activity reveals hallmarks of an optimal internal model of the environment},
  {\em Science} {\bf 331} (2011), no.~6013 83--7.

\bibitem{sadacca2012sodium}
B.~F. Sadacca, J.~T. Rothwax, and D.~B. Katz, {\it Sodium concentration coding
  gives way to evaluative coding in cortex and amygdala},  {\em The Journal of
  Neuroscience} {\bf 32} (2012), no.~29 9999--10011.

\bibitem{yamamoto1981cortical}
T.~Yamamoto, N.~Yuyama, and Y.~Kawamura, {\it Cortical neurons responding to
  tactile, thermal and taste stimulations of the rat's tongue},  {\em Brain
  research} {\bf 221} (1981), no.~1 202--206.

\bibitem{yamamoto1988sensory}
T.~Yamamoto, R.~Matsuo, Y.~Kiyomitsu, and R.~Kitamura, {\it Sensory inputs from
  the oral region to the cerebral cortex in behaving rats: an analysis of unit
  responses in cortical somatosensory and taste areas during ingestive
  behavior},  {\em Journal of neurophysiology} {\bf 60} (1988), no.~4
  1303--1321.

\bibitem{maier2013neural}
J.~X. Maier and D.~B. Katz, {\it Neural dynamics in response to binary taste
  mixtures},  {\em Journal of neurophysiology} {\bf 109} (2013), no.~8
  2108--2117.

\bibitem{grossman2008learning}
S.~E. Grossman, A.~Fontanini, J.~S. Wieskopf, and D.~B. Katz, {\it
  Learning-related plasticity of temporal coding in simultaneously recorded
  amygdala--cortical ensembles},  {\em The Journal of neuroscience} {\bf 28}
  (2008), no.~11 2864--2873.

\bibitem{bermudez2014forgotten}
F.~Bermudez-Rattoni, {\it The forgotten insular cortex: its role on recognition
  memory formation},  {\em Neurobiology of learning and memory} {\bf 109}
  (2014) 207--216.

\bibitem{de2006neural}
I.~E. de~Araujo, R.~Gutierrez, A.~J. Oliveira-Maia, A.~Pereira, M.~A.
  Nicolelis, and S.~A. Simon, {\it Neural ensemble coding of satiety states},
  {\em Neuron} {\bf 51} (2006), no.~4 483--494.

\end{thebibliography}\endgroup
\bibliographystyle{JHEP}

\end{document}